\pdfoutput=1
\documentclass[%
reprint,
 amsmath,amssymb,
 aps,
 onecolumn,
notitlepage,10pt]{revtex4-1}
\usepackage{color}
\usepackage{chemarr}
\usepackage{epsf,mathtools}
\usepackage{graphicx}
\usepackage{dcolumn}
\usepackage{bm}
\usepackage{subfig}
\usepackage{caption}
\usepackage{epstopdf}
\newcommand{\us}{\underline{s}}
\newcommand{\PF}{RR }
\newcommand{\MSA}{VSD }

\begin{document}
\title{Identifying relevant positions in proteins by Critical Variable Selection}
\author{Silvia Grigolon $^{1,*}$, Silvio Franz $^{2,\dagger}$, Matteo Marsili $^{3,\ddagger}$}
\affiliation{$^1$ London Research Institute, The Francis Crick Institute, 44, Lincoln's Inn Fields, London WC2A 3LY, United Kingdom, \\
$^2$ LPTMS - Laboratoire de Physique Th\'{e}orique et Mod\`{e}les Statistiques, UMR CNRS 8626, Univ. Paris-Sud, Universit\'e Paris-Saclay, 91405 Orsay, France,\\
$^3$ Abdus Salam International Center of Theoretical Physics, Strada Costiera 11, 34151 Trieste, Italy}
%
\email{silvia.grigolon@gmail.com}

\email{$^{\dagger}$ silvio.franz@lptms.u-psud.fr}

\email{$^{\ddagger}$  marsili@ictp.it}

\begin{abstract}

Evolution in its course found a variety of solutions to the same optimisation problem. The advent of high-throughput genomic sequencing has made available extensive data from which, in principle, one can infer the underlying structure on which biological functions rely. 


In this paper, we present a new method aimed at extracting sites encoding structural and functional properties from a set of protein primary sequences, namely a Multiple Sequence Alignment. The method, called Critical Variable Selection, is based on the idea that subsets of relevant sites correspond to subsequences that occur with a particularly broad frequency distribution in the dataset. By applying this algorithm to \emph{in silico} sequences, to the Response Regulator Receiver and to the Voltage Sensor Domain of Ion Channels, 
we show that this procedure recovers not only information encoded in single site statistics and pairwise correlations but it also captures dependencies going beyond pairwise correlations. 
The method proposed here is complementary to Statistical Coupling Analysis, in that the most relevant sites predicted by the two methods markedly differ. 
We find robust and consistent results for datasets as small as few hundred sequences, that reveal a hidden hierarchy of sites that is consistent with present knowledge on biologically relevant sites and evolutionary dynamics.  
This suggests that Critical Variable Selection is able to identify in a Multiple Sequence Alignment a \emph{core} of sites encoding functional and structural information.

\end{abstract}

\maketitle

\section{Introduction}

The structure and function that proteins perform inside cells is encoded in their amino acid sequence \cite{biochembook}. Yet sequences are subject to biological evolution, hence the same protein or protein domain may correspond to remarkably different sequences for organisms that are far apart on the evolutionary tree. 
What constrains evolution is precisely the requirement that the structure and biological function be conserved \cite{overington1990, fodor2004, socolich2005}. The sequence of a given protein in different organisms can be regarded as a collection of ``solutions'' that evolution has found to the same optimisation problem. This observation lies at the basis of methods for inferring the way in which structure and functions are encoded in the sequences of amino acids across different species \cite{lunt2010}. The first step consists in compiling a  database of sequences for  a given protein (domain) across species, called Multiple Sequence Alignment (MSA)\footnote{Alignment is necessary in order to compare different positions along the sequence. Here we take the alignment as given, avoiding to enter into the intricacies of algorithms for multiple sequence alignment.}. 
Secondly, by {\em decoding} the statistical traces that constrained evolution leaves in the MSA allows one to {\em reverse engineering} those positions that play a relevant role.
Thus, the frequency of mutations on single sites reveals those positions along the sequence that are ``protected'' from mutations, either because they are associated to important biological functions or because they are vital for the stability of the tertiary structure. Furthermore, correlations in the mutation of pairs of sites carry information that can be reverse engineered to reveal contacts in the 3D structure \cite{lunt2010,weigt2009,morcos2011,cocco2013, ekeberg2013, chris2014, chris2014b, ekeberg2014}. Coevolution on larger subsets of sites can be spotted by an extension of Principal Component Analysis, called Statistical Coupling Analysis (SCA)\cite{halabi2009} , that aims at identifying regions that are associated to  functional domains \cite{halabi2009, cocco2013}.

Yet, in all these examples, inference techniques are limited by and rooted in statistics that do not go beyond pairwise correlations. Indeed, available data barely allows one to estimate pairwise correlations, not to speak of higher order statistics. 
Yet, there is no reason why evolution should use only pairwise correlations to encode biological functions in amino acid sequences. As a matter of fact, selection operates on the whole sequence. 

In this paper, we propose a new statistical non-parametric method going beyond pairwise correlations for the analysis of MSAs of a given protein (domain) family. The method, that we call {\em Critical Variable Selection} (CVS), is based on the conclusions of Ref. \cite{MMR} that sampling relevant degrees of freedom of a complex system generates datasets with broad distribution of frequencies \footnote{{\em Critical}, besides referring in general to an inquisitive attitude, refers to the fact that broad frequency distributions, in particular power law distributions, have been associated with critical phenomena in statistical physics. In this respect, Ref. \cite{MMR} argues that samples probing relevant variables  ``look critical''.}. This conclusion is further refined in Ref. \cite{AHMM} within a Bayesian model selection approach. This paper is the first systematic attempt to exploit these ideas for identifying the subset of relevant positions in a biological dataset. Unlike other methods, CVS aims at distinguishing relevant from irrelevant variables, on purely information theoretic grounds, {\em without assuming an a-priori}  criterion 
 of relevance. The aim of our method is to unveil first of all whether there is a well-defined hierarchy in a given set of variables and, as a consequence, to characterise it. Here, relevance becomes a sample specific concept, depending on the way the sample has been assembled. As an example, if no hierarchy is present among the variables, we would not expect CVS to identify a distinction between relevant and irrelevant. In this sense, our method differs from methods such as Direct Coupling Analysis (DCA)  \cite{morcos2011} aimed at identifying specific features (e.g., contacts between amino acids), that call for models (with pairwise interactions) in which those features are related to sufficient statistics. The downside of our approach is that, while it is easy to validate methods that target a specific goal on hundreds of protein families (e.g., by comparing DCA predictions with measured inter-residues distances), the validation of our method requires in-depth analysis of the protein family, because what is relevant in one family needs not be relevant in another. 
Therefore, our analysis will focus on two specific families, the Response Regulator Receiver (\PF, PFAM ID PF00072) and the Voltage Sensor Domain of the Ion Channels (\MSA, PFAM PF00520), though it has been performed on several other families. The sequences in these  two families have  nearly the same length, but the size of the database differs by an order of magnitude. This allows us to probe CVS as the depth of the dataset varies, which is an important dimension. 

The paper is structured as follows. After a general introduction on the method's formulation and the related algorithm, we show the outcome of CVS when applied to an \textit{in silico} sequence, a paradigmatic example to understand what kind of results such method affords. We then proceed with the study of real biological sequences, showing that CVS can  consistently identify the existence of an underlying ranking in relevance among sites on actual samples. These preliminary sections are then followed by more in depth analysis, showing the ability of CVS to unveil information going beyond pairwise correlations and the robustness of the method with respect to sampling and evolutionary biases. We then compare our results with the prediction of state-of-the-art methods, such as single-site conservation, Statistical Coupling Analysis (SCA) \cite{halabi2009} and Direct Coupling Analysis (DCA) \cite{morcos2011, lunt2010, weigt2009}.
Our whole analysis shows that in the cases studied, CVS identifies a core of interdependent positions that is denser and tighter than that identified by SCA. Finally, we discuss the biological relevance of the site identified by CVS in the Response Regulator Receivers and the Voltage Sensor Domain of the Ion Channels, in relation to site conservation, functional sites and solvent accessible surface scores.     

\section{Critical Variable Selection}

\subsection{Theory}


Let us consider a Multiple Sequence Alignment of homologous proteins composed of $M$ sequences of length $L$, $\vec{s}^{\alpha}=(a_1^\alpha,\ldots,a_L^\alpha)$, where $\alpha=1,...,M$ labels the sequence and $a_i^\alpha$ is the $i^{\rm th}$ amino acid of the $\alpha^{\rm th}$ sequence\footnote{The values that $a_i^\alpha$ can take correspond to the different amino acids. Including gaps and ambiguous amino acids, $a_i^\alpha$ can take up to 21 values in the examples studied in this paper.}. 
Each sequence $\vec{s}=(a_1,\ldots,a_L)$ can be thought of as a ``solution'' of how the same biological function is achieved in the specific environment where that particular sequence has evolved. Following Ref. \cite{MMR}, we think of this as an optimisation problem of an unknown function 
that also depends on unobserved variables. Sub-sequences that occur very often are optimal under broader conditions with respect to sub-sequences that occur rarely. This suggests that the frequency with which a given sub-sequence occurs provides a noisy estimate of the function being optimised. So highly conserved sites are expected to be functionally relevant, whereas sites with high variability are unlikely to be relevant.
This suggests to look for subsets of variables such that the frequency with which the corresponding subsequences occur has a larger variability in the MSA. Loosely speaking, these are relevant because they ``probe'' steeper gradients of the function being optimised. Conversely, variables that respond to the details of the cellular environment or that are subject to random mutations will instead generate incoherent changes in the frequency distribution. This idea is best illustrated with a specific example, referring interested readers to Refs. \cite{MMR,AHMM} for a more detailed theoretical discussion.

\begin{figure*}[ht] 
\centering                              
\includegraphics[scale=0.7]{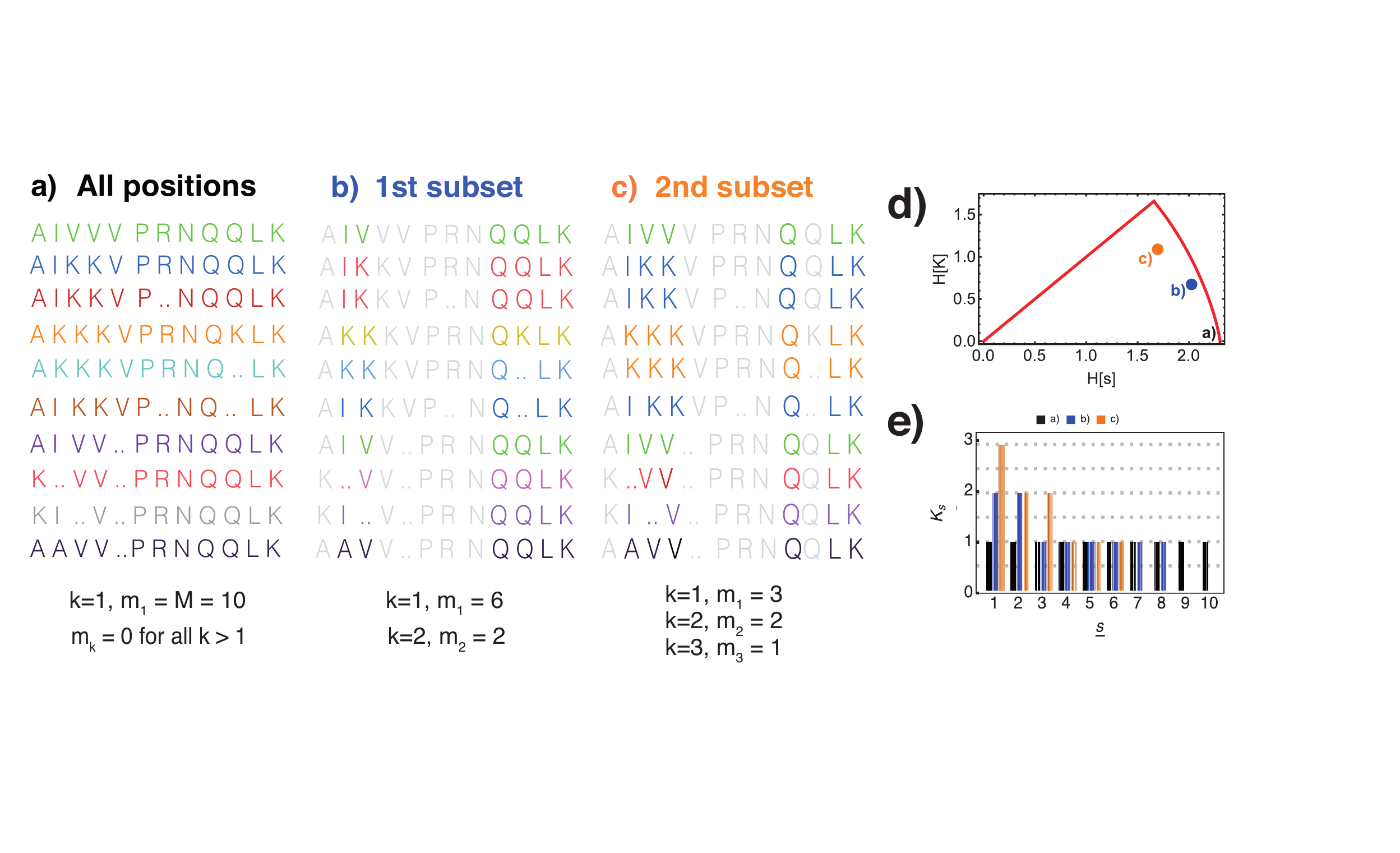}
\caption{a), b) and c) Example of three samples obtained from the same one when considering different positions, highlighting the corresponding different states, $\us$. Counts $k$ and their multiplicities $m_k$ are shown as well below the three panels. d) Relevance $H[K]$ as a function of the resolution, $H[\vec s]$, for the three samples represented in a), b) and c) below the red line depicting the Poissonian case. e)  Distributions of the counts, $K_{\us}$, for the different states $\us$ observed in the three samples in a), b) and c).}
\label{fig:hk_general}
\end{figure*}

Let the MSA in Fig. \ref{fig:hk_general} a) be a compilation of sequences across the evolutionary tree for protein domains that perform the same biological function. Notice that each sequence $\vec s^\alpha$ is unique, which reflects the fact that the MSA should sample as uniformly as possible the evolutionary process. Let us assume that, of the $L=12$ positions, $n=6$ are relevant for the function of interest and the remaining ones are irrelevant. Fig. \ref{fig:hk_general} b,c) report two subsets $I_b=\{2,3,9,10,11,12\}$ and $I_c=\{2,3,4,9,11,12\}$ of positions. For each $I\subseteq \{1,2,\ldots,L\}$ we split the sequence $\vec s=(\us_I,\bar s_I)$ in the subsequence $\us_I=\{a_i:~i\in I\}$ over the (putatively) relevant sites and in the subsequence $\bar s_I$ of the remaining sites. Let 
\[
k_I(\us)=\sum_{\alpha=1}^M\delta_{\us,\us_I^\alpha}
\]
be the frequency with which different subsequences $\us_I$ occur in the sample. As Fig. \ref{fig:hk_general} shows, different choices of $I$ lead to different distribution of frequencies, that in general differ from the flat frequency distribution $k(\vec s)=1$ corresponding to the entire sequence $\vec s$ (black histogram in Fig. \ref{fig:hk_general}e). The key observation in Ref. \cite{MMR} is that a broader distribution of frequency (the one corresponding to $I_c$ here) will be more informative of the function being optimised, because it allows to distinguish more states in terms of their ``functionality''. 
A quantitative measure of relevance is then given by the entropy of the frequency distribution 
\begin{equation}
\label{ }
H[K_I]=-\sum_{k}\frac{km_I(k)}{M}\log \frac{km_I(k)}{M}
\end{equation}
where 
\begin{equation}
\label{ }
m_I(k)=\sum_{\us}\delta_{k,k_I({\us})}
\end{equation}
is the number of subsequences $\us_I$ that occur $k$ times in the MSA. 
$H[K_I]$ is a proxy for the (log of the) number of states that the subset $I$ allows one to resolve in frequency. 
Notice that $H[K_I]$ is different from the entropy of the sequence $\us_I$
\begin{equation}
\label{ }
H[\us_I]=-\sum_{\us}\frac{k_I(\us)}{M}\log\frac{k_I(\us)}{M}=-\sum_{k} \frac{k m_I(k)}{M}\log  \frac{k}{M}.
\end{equation}
While $H[\us_I]$ measures {\em resolution}, $H[K_I]$ measures {\em relevance}. Ref. \cite{AHMM}, to which we refer, gives compelling arguments for this conclusions on the basis of a Bayesian model selection framework, showing that $H[K]$ correlates with the number of parameters that can be estimated on the basis of the data. In this respect, finding the subset $I$ that has maximal $H[K_I]$ is tantamount to finding those variables that are described by the richest model. 

Going back to our example, Fig. \ref{fig:hk_general} d) shows the values of $H[\us]$ and $H[K]$ for the frequency distributions a), b) and c) in the figure (we will suppress the index $I$ if that creates no ambiguity in what follows). The solid line provides an upper bound to the relation between $H[\us]$ and $H[K]$ \cite{MMR}. Notice that the inclusion of totally conserved sites, such as $\{9,11,12\}$ does not affect the value of either $H[\us]$ and $H[K]$, whereas inclusion of position $4$, that covaries with $2$ and $3$, engenders a smaller decrease in $H[K]$ with respect to position $10$. 

In general, we shall posit that the relevance of a sub-set of positions $I$ is quantified by $H[K_I]$. The aim is thus to find the sub-set $I$ of not necessarily contiguous positions that maximises $H[K_I]$. As in the example above, we shall look for maxima over sub-set $I$ of $n=|I|$ positions:
\begin{equation}
\label{CVS}
I_n^*={\rm arg}\max_{I:|I|=n}H[K_I].
\end{equation}
The prediction of relevant sites based on the maximisation of $H[K]$ will be called  {\em Critical Variable Selection} (CVS) in what follows. 
The algorithm that searches for solutions of this problem is described below. Before then, it is worth to remark few important points.

First, relevance is a relative concept. Here it is relative to the criterium with which the sequences of the MSA have been selected. The quality of the results depends ultimately on the quality of the data and on the algorithms for the Multiple Sequence Alignment that is used. We shall not discuss issues related to MSA algorithms and rely on MSA curated and compiled by others \cite{cocco2013,MSApaper} that we regard as benchmarks.

At odds with other methods \cite{halabi2009,weigt2009}, CVS does not rely on an explicit definition of correlation between sites. Indeed, it is based on statistics (the frequency) that goes beyond pairwise correlation and it uncovers information that goes beyond pairwise correlations.

Furthermore, this method assumes as only parameter the number $n$ of positions that sets the resolution in the optimisation problem. 
Changing $n$ allows us to see how the set of relevant sites expands. If the MSA contains a hidden hierarchy of relevant sites, we expect that varying $n$ is equivalent to ``zooming'' in and out on the subset of relevant sites, thereby revealing the hierarchy. We also expect this hierarchy to be corrupted by noise, when $n$ becomes too large. So the parameter $n$ will be used for a mere aid in the search for a hierarchy of relevant sites. 

\subsection{The algorithm}

Following our previous considerations, in this section we want to propose an operative way of identifying the subsets of $n$ positions, $I^{\ast}(n)$, maximising the relevance function $H[K]$. In order to implement Eq. (\ref{CVS}), we employ a simple greedy gradient ascent algorithm: starting from a random choice of the subset $I$ of $n$ positions, we iterate the following steps:
\begin{enumerate}
\item Construct a new tentative subset $I'=(I/\{i\})\cup \{i'\}$ by 
changing one position $i\in I$, chosen at random, into a randomly chosen position $i'\not\in I$;
\item if $H[K_{I'}]\ge H[K_I]$, accept the move and $I\to I'$ otherwise leave $I$ unchanged. 
\end{enumerate}

The algorithm stops when $ H[K_I]$ does not change for a sufficiently large number of (attempted) moves \footnote{In the examples studied in this paper, $20L$ steps were typically found to be enough for this condition to be met.}. 
Typically, for a given value of $n$, the algorithm does not produce a single maximum but rather a population of local maxima with similar value of $H[K]$. In order to get a consistent statistics and fully explore such maxima profiles, we 
run the algorithm $R$ times for each value of $n$ starting from randomly chosen subsets $I$ (typically $R \simeq 100\div 1000$). Therefore the algorithm returns a distribution of solutions, each corresponding to a local maximum of $H[K_I]$. 
In order to assess the \emph{relevance} of each position, we count the number $c_i(n)$ of times that position $i$ is selected in the $R$ solutions. We shall call this value simply the \emph{count}. Upon running the algorithm on different values of $n$, each position can be assigned a \emph{total count}, given by $C_i=\sum_{n} c_i(n)$, which provides information on the ``relevance'' landscape of a given protein family.   

\section{The data}

In order to understand the typical output of our method, we applied it to three different datasets, namely one \emph{in silico} family of sequences and two biological MSAs. The former is made of $M=10^4$ sequences of length $L=64$ where, for simplicity, each amino acid $a^{\alpha}_i$, $i=1,...,L$, $\alpha=1,...,M$ can take only two different values, i.e., either 0 or 1.
Each sequence in the dataset is generated in order to include sites with a different degree of conservation and of mutual dependence. Each generated sequence is divided into four regions:
\begin{itemize}
\item a \emph{core} made of 5 highly correlated positions $a_i^\alpha$, $i=1,...,5$. In order to create non-trivial correlations that go beyond second-order statistics, these values are assigned taking the first five bits in the binary representation of a random number $X\in [0,1]$ drawn from the pdf $p(x)=1/(2\sqrt{x})$;
\item a set of 12 \emph{subordinated sites} that take values that are noisy functions of the core variables. For $i=6,...,15$, defining $s_i^\alpha=2a_i^\alpha-1$, 
\[
P\{s^{\alpha}_i=s^{\alpha}_k s^{\alpha}_{l} s^{\alpha}_{n}\}=0.95,
\]
where $1\le k<l<n\le 5$ take all possible combinations, whereas 
$s^{\alpha}_{16}$ and $s^{\alpha}_{17}$ take values $s^{\alpha}_{16}=s^{\alpha}_1 s^{\alpha}_{2} s^{\alpha}_{3} s^{\alpha}_4$ and $s^{\alpha}_{17}=s^{\alpha}_2 s^{\alpha}_{3} s^{\alpha}_{4} s^{\alpha}_5$ in 95\% of the cases and the opposite values otherwise.

\item a set of highly \emph{conserved sites} made of 17 positions, where $s^{\alpha}_i$ takes the same value in 95\% of the cases.

\item a set of \emph{random sites} made of the remaining 31 positions, obtained by drawing each of them, independently at random with $P\{s^{\alpha}_i=0\}=0.5$.
\end{itemize}
An example of this dataset is shown in Fig.\ref{fig:marsilina} a).

The biological MSAs analysed in the following refer instead to two different protein families, i.e., the Response Regulator Receiver and the Voltage Sensor domain of the Ion Channels.

Response Regulator receivers (RR, PFAM ID PF00072) are part of two-components signal transduction machineries allowing cells to sense and respond to a high variety of environments. These two components are usually made of a histidine kinase (HK) aimed at sensing the surrounding environment, i.e., controlling the input, whose signal is received by the response receiver domain (RR), that in turn triggers cell physiology and response \cite{west2001}. The RR consists of other two subdomain, a N-terminal response regulator receiver domain and a variable C-term domain aimed at DNA binding. RRs adapted to a wide variety of signals and their evolutionary spread as well as their availability makes this system suitable for any statistical analysis, and in particular for our technical purposes. Within the context of our analysis, we focused on the dataset used in \cite{cocco2013}, made of $N=62074$ sequences of length $L=112$. 

Voltage-dependent ion channels are biomolecular machines aimed at measuring changes in the cell transmembrane voltage and, because of their ubiquity, are found to be as well connected to many heritable diseases \cite{catterall2010}. Structurally, these channels are made of four identical subunits, the Voltage Sensors, each of them divided in turn into six segments (S1-S6). Hereby, we are going to focus on the first four segments building up the so-called Voltage-Sensor Domain (VSD), (S1-S4), found to undergo conformational changes during voltage sensing. 
Our dataset is made of $M=6652$ sequences of length $L=114$, already curated in \cite{MSApaper}. Note that the size of this dataset is much lower than that of the Response Regulator Receiver, although the length of the sequences is similar. This sample can be then informative to test the ability of our method to extract relevant information even in the strong undersampling regime.

\section{Typical behaviour of the algorithm}

In this Section we illustrate the behaviour of the CVS algorithm, by applying it to  synthetic and biological MSA.

\subsection{\emph{In silico} sequences}

\begin{figure*}
\centering
\includegraphics[scale=0.6]{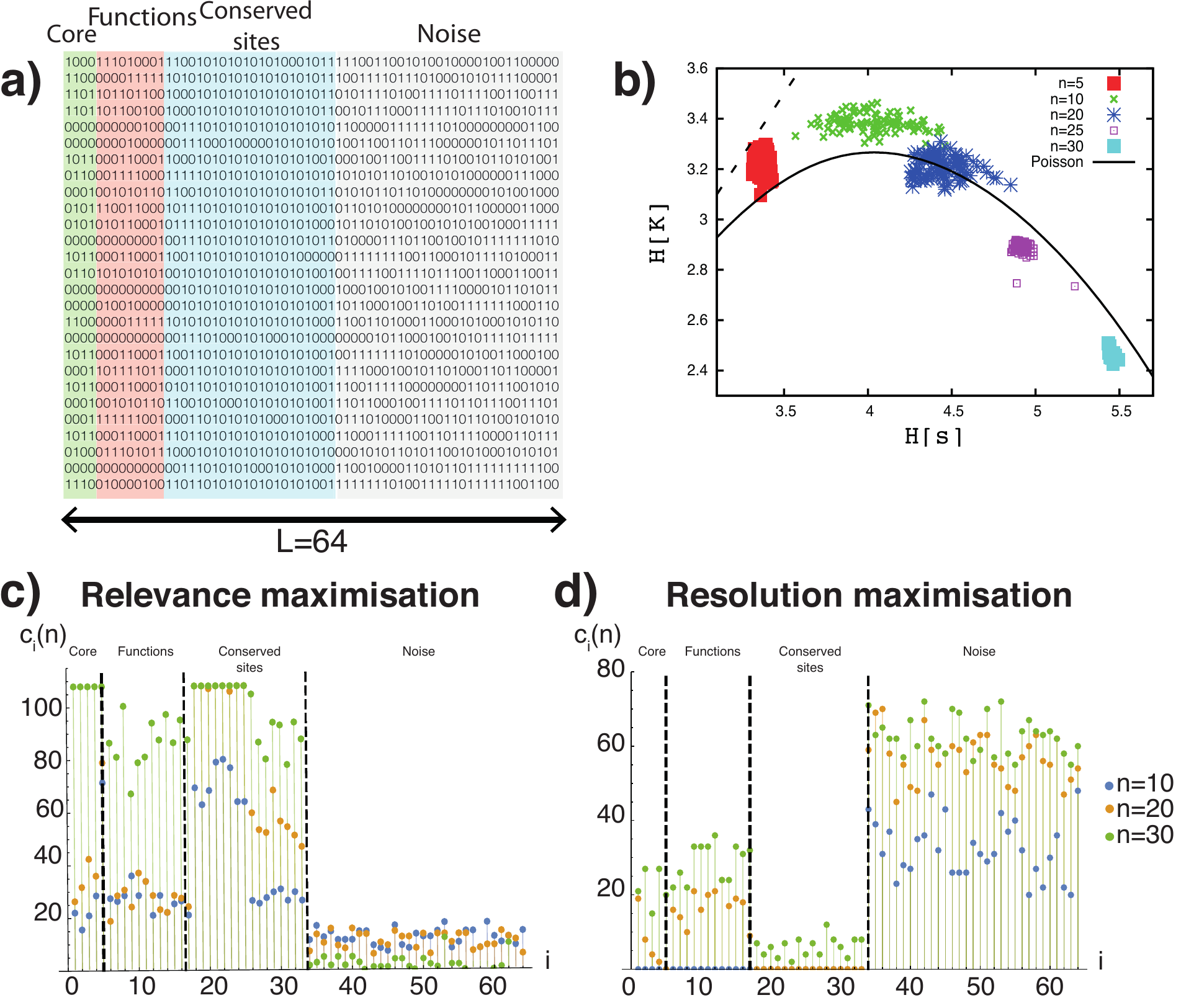}
\caption{a) Example of a typical sample of the \emph{in silico} sequences used to test Critical Variable Selection. The different parts constituting each sequence are highlighted as described in the Text. b) Relevance, $H[K]$, as a function of the resolution, $H[\us]$, for the sites maximising the relevance varying the subsequence length $n$. The results for Poissonian-distribuited multiplicities $m_k$ are shown as well \cite{MMR}. c) Single site counts for different subsequence lengths, $c_i(n)$, obtained by maximising the relevance, $H[K]$. The results here obtained are deeply different than those shown in d) as the most frequent selected positions are indeed the ones belonging to the core and the conserved sites. d) Single site counts for different subsequence lengths, $c_i(n)$, obtained by maximising the resolution, $H[\us]$. As described in the Text, such a quantity tends to select by definition those positions showing maximal variability across the dataset.}
\label{fig:marsilina}
\end{figure*}

Let us now focus on the typical behaviour and the outcomes of this algorithm when run on the \emph{in silico} sequence  dataset built \emph{ad hoc} to understand the typical output of Critical Variable Selection. Fig. \ref{fig:marsilina} a) shows a sample of the dataset. As described before, besides highly conserved sites, the synthetic MSA also contains sites with non-trivial correlation, going beyond second order statistics. We run the CVS algorithm 100 times for $n=5,10,15,20,25$ and $30$. 
Fig. \ref{fig:marsilina} b) shows the values of $H[\us]$ and $H[K]$ obtained in each run, for each value of the sub-sequence length $n$. With increasing $n$, CVS returns subsets of sites with higher and higher resolution $H[\us]$. At the same time, $H[K]$ exhibits a non-monotonic behaviour as it first increases from $n=5$ to $n=10$ and then it decreases. 
Close to the  maximum, the points corresponding to different runs are scattered over a large region, suggesting that CVS prediction is most noisy. Indeed, the optimal value of $n$ does not correspond to the maximum of $H[K]$ because there $n$ is too small to capture adequately the statistical dependences in the MSA. Much insight is provided by looking at how the sample of optimal sub-sets of positions evolve as $n$ varies. Fig. \ref{fig:marsilina} c) shows the counts $c_i(n)$ for each position, for $n=10,20$ and $30$. CVS first identifies highly conserved sites, together with those core sites that are most conserved. Next, core sites and the subordinated sites are selected when $n$ increases. Notice that the separation between relevant and irrelevant sites becomes sharper and sharper as $n$ increases: while the counts of relevant sites increase with $n$, those of random sites decrease as $n$ increases. This fact implies that sites that are found relevant by CVS at a given $n$ remain relevant when $n$ increases. This is a strong indication that CVS is uncovering a hidden hierarchy of relevant positions. This hierarchy is revealed by the ranking of positions in terms of total counts $C_i$. The top ranked sites turn out to be the highly conserved ones, followed by the core sites and the functional sites. In this case, the noisy part is very poorly ranked.

Fig. \ref{fig:marsilina} d) contrasts these results with what one obtains by maximising the resolution $H[\us]$ instead. This procedure is expected to select highly variable sites. Indeed, we observe the opposite scenario, where counts $c_i(n)$ increase with $n$ for the random sites and decrease with $n$ for conserved and functional sites. 

In summary, as $n$ increases, CVS reveals the hierarchy of conservation and dependence that is hidden in the MSA. In order to reveal this structure it is important to run CVS for increasing values of $n$, at least as long as the separation between relevant and irrelevant sites remains sharp.

\subsection{Biological sequences: sample size, reshuffling and reweighing}

\begin{figure*}[htbp]
\centering
\includegraphics[scale=0.4]{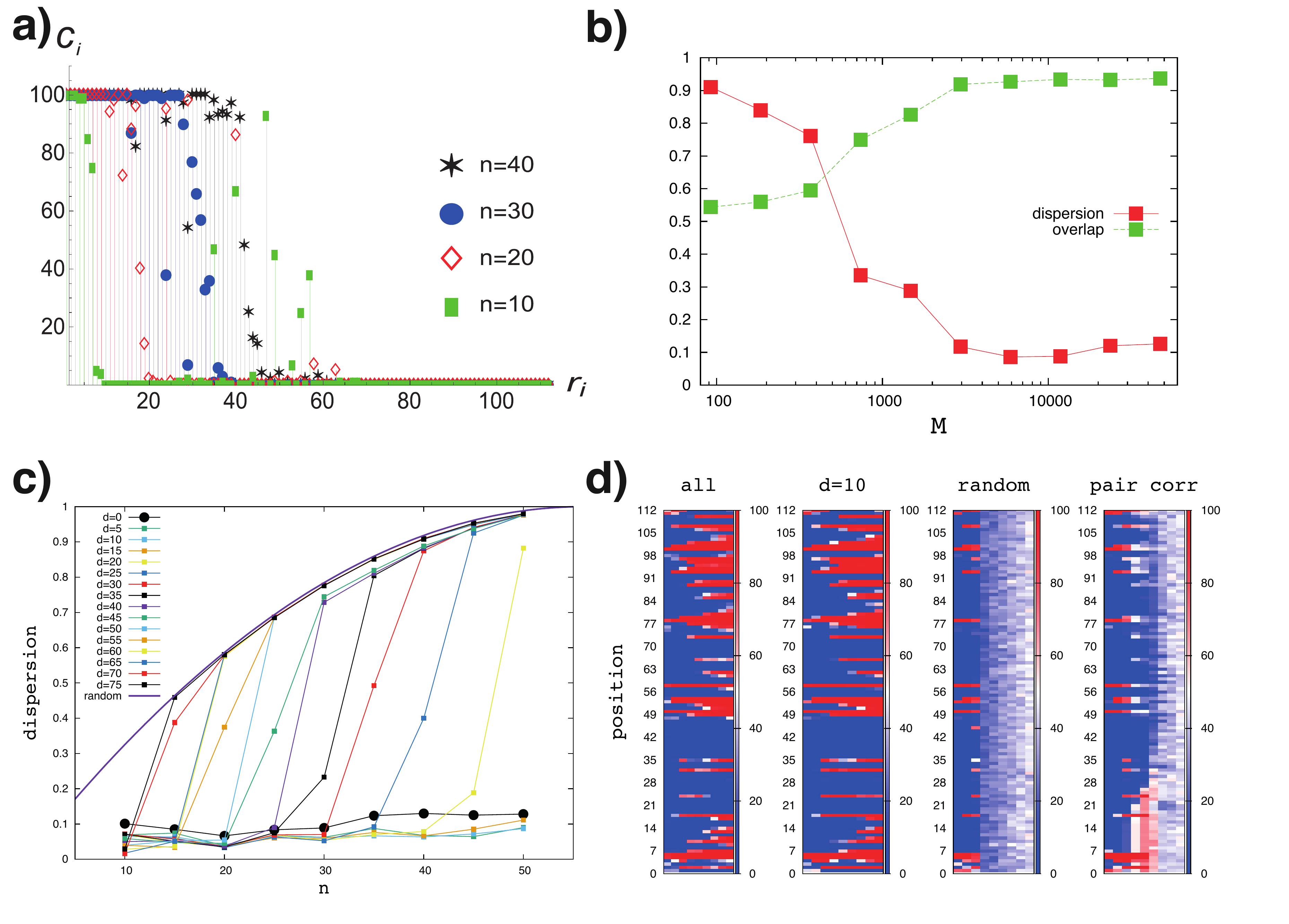}
\caption{a) Critical Variable Selection outcome for the Response Regulator Receiver dataset: single-site counts, $c_i(n)$ as a function of their rank $r_i$ with respect to single-site total counts $C_i$, for $n=10,20,30$ and $40$ ($R=100$). 
Different colours refer to different subsequences' lengths as shown in the side legend. b) Overlap and dispersion of the CVS outcome on datasets built up taking into account $M$ sequences of the actual one (legend shown as inset). c) Dispersion between CVS outcome on different realisations of the dataset and the randomly picked sites as a function of the subsequence length $n$. Here the different realisations correspond to the actual dataset collapsed by using a similarity threshold equal to $d$ amino acids (legend shown as inset). d) Heat maps showing the single-site counts, $c_i(n)$, as a function of the subsequence length $n=10,15,\ldots,50$ (x-axis) and the position along the sequence (y-axis) for four different realisations of the dataset, i.e., the actual one, the one where sequences are collapsed by a similarity threshold $d=10$ amino acids, the reshuffled dataset constraining single-site frequencies and the reshuffled dataset constraining pairwise correlations. }
\label{fig:PF72technical}
\end{figure*}

While in the study of in-silico MSA the ground truth of the generating algorithm is available for the validation of results, for real biological sequences one has to resort to different information, such as annotations of known functional sites or structural properties. 

Before doing that, it is worth to discuss the performance of CVS with respect to important aspects of the MSA: {\em i)} sample size $M$, {\em ii)} evolutionary bias and {\em iii)} the nature of correlations. 



We illustrate these aspects for the MSA of the RR domain. The analysis presented here can be considered as a preliminary study that can be applied to any MSA in order to asses the statistical robustness of the output of CVS.

The emergence of a hierarchy of relevant sites can be spotted by ranking sites according to their total count $C_i$ and plotting the counts $c_i(n)$ as a function of their rank, for different values of $n$, as in Fig.\ref{fig:PF72technical} a). As a quantitative measure of the sharpness of the separation between relevant and irrelevant sites, we take the \emph{dispersion}, which is defined as
\begin{equation}
\mbox{dispersion}=\frac{1}{L} \sum_i 4 p_i (1-p_i)
\end{equation}
where $p_i = \frac{c_i}{R}$ is the probability of site $i$ to be selected by CVS, estimated as the fraction of times position $i$ is selected in the $R$ independent runs. 
A small value of the dispersion implies that all sites are either selected in most runs ($p_i\simeq 1$) or rarely selected ($p_i\simeq 0$). As a benchmark, in the random case we expect counts to be $c_i\simeq n/L$, so that when CVS is purely dominated by noise the dispersion is given by $\approx 4(n/L)(1-n/L)$. 

We use this measure to test the robustness of CVS with respect to the sample size $M$. Starting from the original dataset for \PF with poorly gapped sequences ($M_0= 47349$ and number of gaps for each sequence less than 4 $\%$) we iteratively reduce $M$ by a factor of two, by randomly selecting a subset of half of the sequences each time. Fig.\ref{fig:PF72technical} b) shows the behaviour of the dispersion, for $n=40$, as a function of the sample size $M$. For $M\approx 100$ this approaches the random limit $\approx 0.92$, whereas for $M>600$ a well defined structure emerges, which becomes very robust for $M>2000$. 

A second useful measure is the overlap  
\begin{equation}
\mbox{overlap}=\frac{1}{L} \sum_i \Big[ p^0_i p_i^1+(1-p^0_i)(1-p_i^1) \Big],
\end{equation}
between the outputs $p_i^0$ and $p_i^1$ of CVS for two different datasets. This has the simple interpretation as the probability, averaged over sites, that CVS predicts the same relevance for a site $i$ in the two 
datasets.
In Fig.\ref{fig:PF72technical} b) $p_i^0$ refers to CVS for the full dataset whereas $p_i^1$ refers to the reduced dataset with $M$ sequences. In the random case, we expect an overlap $\approx 0.54$, which is the value we observe for $M\simeq 100$. For larger values of $M$ we see that the overlap converges to values close to one for $M>2000$. In summary, for $M>2000$ we observe a well defined structure of relevant sites, and this structure is the same up to $M=47349$.

A second aspect that requires particular attention is the bias in the datasets that
comes from the fact that sequences belonging to better studied organisms typically occur with enhanced frequencies. In order to partially correct for this bias, we limit our analysis to MSAs of sequences that differ on more than $d$ positions\footnote{These reduced MSAs are obtained by picking sequentially at random the sequences of the original MSA and by including them in the new MSA if their minimal distance to the sequences already included is more than $d$. We checked that the outcomes are robust, in the sense that they do not depend on the specific realisation of this reweighing process.}. Fig.\ref{fig:PF72technical} c) shows the dispersion for different realisations of the dataset as a function of the length of the subsequence, $n$, whereas Fig.\ref{fig:PF72technical} d) shows a density plot of the counts $c_i(n)$ as a function of the position $i$ and of $n$, for the cases $d=0$ and $d=10$ (first two panels). 
The separation between relevant and irrelevant sites predicted by CVS turns out to be sharper for $d=5,10$ and $15$ than in the original dataset (smaller dispersion). This confirms that biases in the sampling of the evolutionary process indeed mask statistical dependences. For larger values of $d$, important traces of the evolutionary dynamics are lost, as suggested by the fact that the dispersion for large $n$ converges to the random limit (e.g. for $d=30$ and $n\ge 40$). This analysis suggests that $d=10$ is a meaningful threshold for the similarity of sequences.

Finally, in order to gauge the nature of the statistical dependencies between positions, we compare the predictions of CVS in the original dataset to those obtained from randomly reshuffled MSAs. 

First, in order to understand how much single site conservation affects the results of CVS, we produced a random MSA where, at each positions, the amino acids were reshuffled across sequences. Secondly, in order to test the relevance of pairwise correlations, we followed the simulated annealing procedure of Ref. \cite{reynolds2013} that allows to produce randomised datasets that preserve pairwise correlations. 

The heat maps of the counts $c_i(n)$ in Fig.\ref{fig:PF72technical} d) show that on the datasets randomised by constraining single site frequencies (third panel) and correlations (fourth panel) CVS is able to infer a hierarchy if $n\leq 15$ and $n \leq 30$ respectively. This translates into the fact that for $n \leq 15$ the structure within the dataset is mainly dominated by single site conservation and then by pairwise correlations up to $n \simeq 30$, generating predictions which are close to random (i.e. $c_i(n)\approx Rn/L$) for $n>30$. By contrast, CVS maintains a low dispersion on the original MSA and on the $d=10$ MSA up to $n=50$ and beyond, highlighting the presence of statistical dependencies in the evolutionary process  going well beyond pairwise correlations. 

Finally, it is also worth noticing that single site conservation does not necessarily imply relevance. Site $112$ in \PF formally appears as
highly conserved in the dataset because it is very often a gap. Interestingly, while this site is picked up as relevant in the reshuffled MSAs, its counts are negligible for $n>20$ in the real dataset.

In summary, this analysis allows us to draw three main conclusions on this specific MSA: (i) Critical Variable Selection provides a robust choice of sites over a wide range of sample size; (ii) Critical Variable Selection provides robust results and selects non-random sites when reweighing sequences with a similarity threshold up to $d=15$ amino acids; (iii) Critical Variable Selection infers a \emph{signal} in the dataset going beyond pure single-site conservation or pairwise correlations.

This analysis can be performed on any MSA corresponding to a given protein domain family, and it provides a preliminary test on whether CVS extracts a robust and non-trivial information from the MSA. The next step is to analyse the biological relevance of the results. But before doing that, we will compare and combine CVS to other methods for the analysis of MSAs.

 \section{Comparison with { different statistical methods:}  Statistical Coupling Analysis { and Direct Coupling Analysis}}
 \label{subsec:comparisonsca}
 
To better understand the behaviour of CVS, we compare its predictions to those
of well-established statistical techniques aimed at identifying relevant sites and sites' relations in proteins: Statistical Coupling Analysis (SCA) \cite{halabi2009} and Direct Coupling Analysis (DCA)\cite{morcos2011}.

Let us start with the comparison with SCA,
referring the interested reader to Ref. \cite{halabi2009} or to the
Supplementary Material for a detailed description. In brief, SCA is based on comparing the pairwise correlation matrix with the one built from a randomly reshuffled MSA with the same single amino acid frequencies at each position. Focusing on those principal components that correspond to eigenvalues that stand out of the distribution of the eigenvalues of the randomised matrix, one can define sectors corresponding to (putative) functional regions (see Supplementary Material). By applying SCA to \PF we identified two relevant principal components.The projections of these components along the sequence allow to represent each amino acid position in a two dimensional plane. The distance of each point from the origin can be taken then as a measure of relevance of the corresponding site, according to SCA. 

\begin{figure}[htbp]
\centering
\includegraphics[scale=0.4]{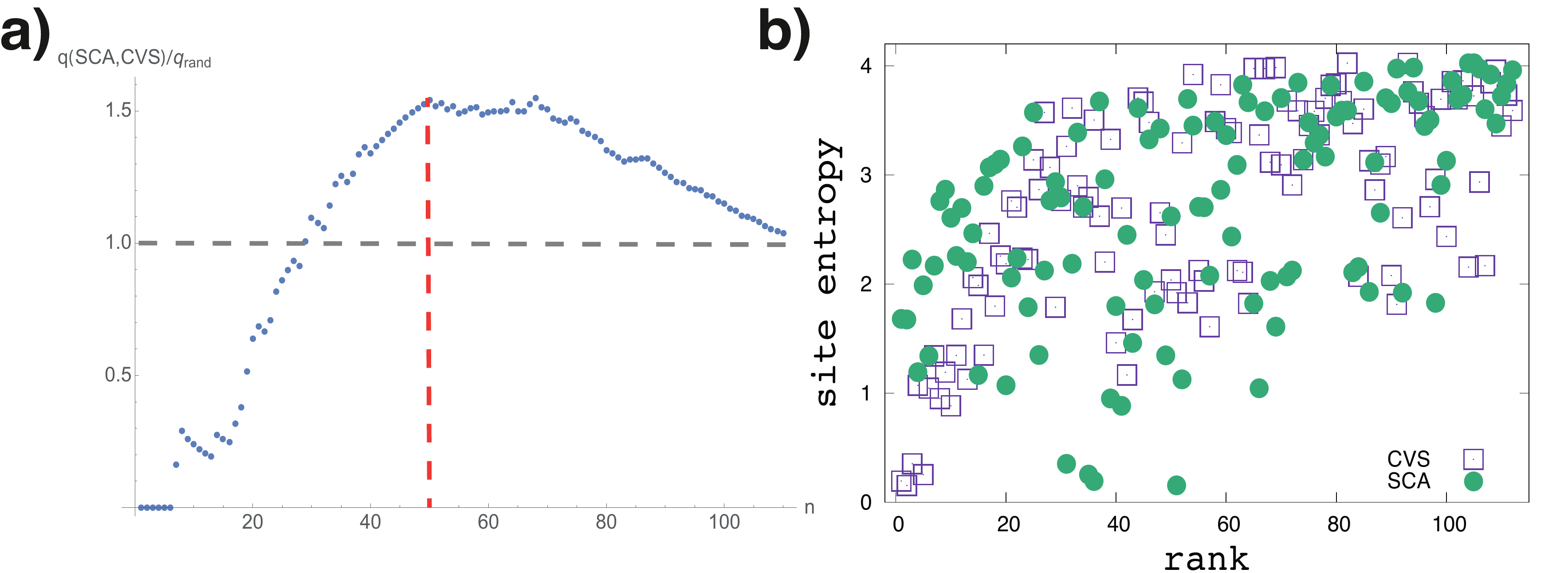}
\caption{a) Overlap between the top $n$ sites obtained by SCA and CVS ($q(SCA,CVS)$) normalised to the random value ($q_{rand}=q_{rand}=(n/L)^2+ (1-n/L)^2$) as a function of the top $n$ sites. For small $n$, this ratio is equal to zero. It increases afterwards, going beyond the random value for $n \simeq 30$ and it tends to $q_{rand}$ for $n \to L$  (dashed grey line). In our comparisons, we took the value of $n$ less than half the sequence ($n \simeq 50$) for which the overlap is maximal (red line). However, one can notice another maximum to be at $n=68$. b) Site entropy as a function of the rank according to CVS (squares) and SCA (circles).}
\label{fig:PF72_lists}
\end{figure} 

\begin{figure*}[h!]
\centering
\includegraphics[scale=0.4]{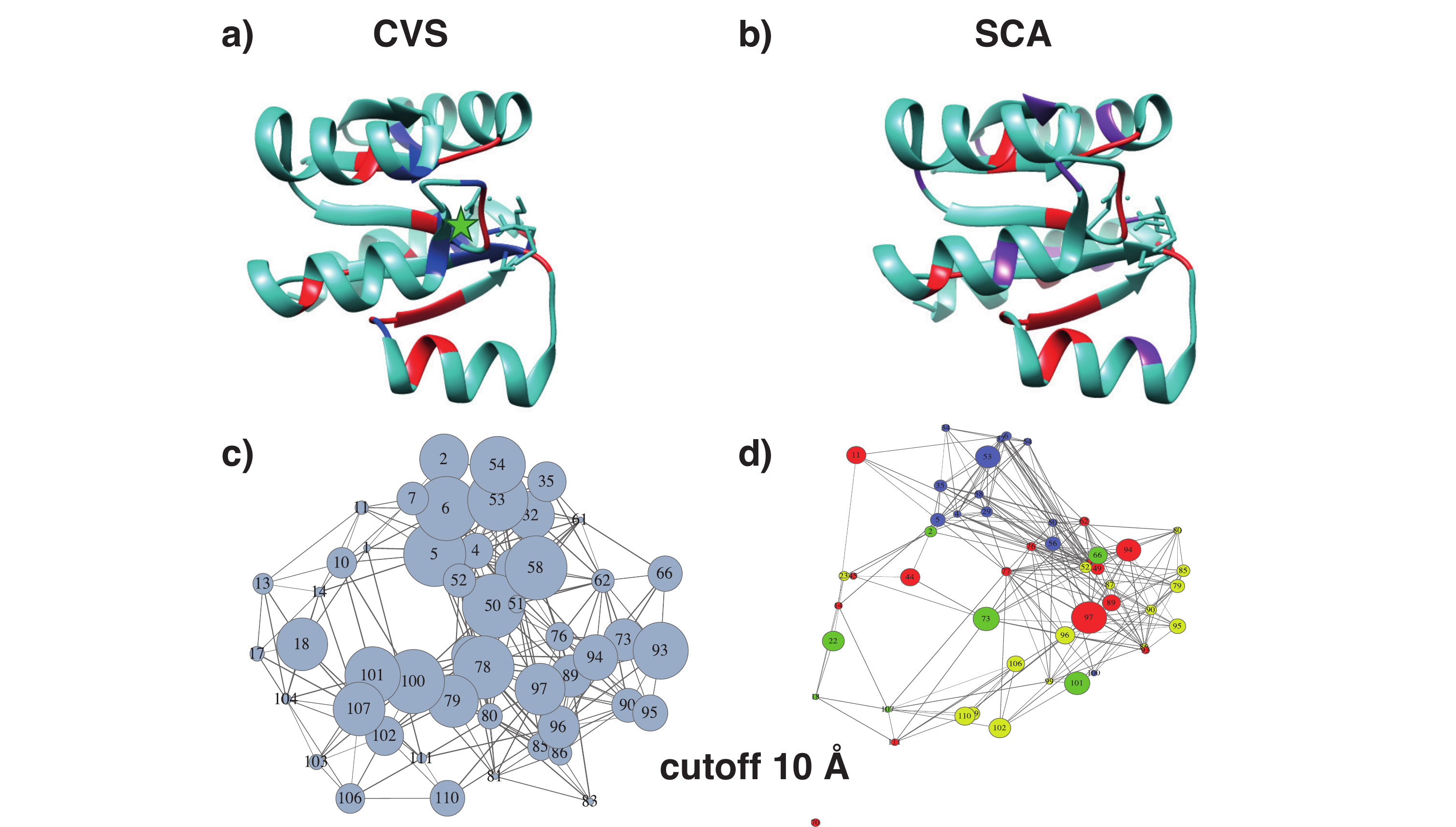}
\caption{a)-b) Top $30$ sites (gapped position excluded) obtained by CVS (a) and SCA (b) visualised on the 3D structure of the response regulator receiver (PDB 1NXW). Common sites are highlighted in red, whereas sites singled out only by CVS (SCA) are marked in blue (purple). The active and phosphorylation site (52 on 1NXW) is marked in green in (a) and selected by CVS among the top 30 sites.
c-d) Network representation of the CVS top 50 sites (c) and of the SCA top 50 sites (d). The size of the nodes is rescaled respectively with the CVS counts $C_i$ and the notion of relevance defined for SCA (i.e., distance from the origin in the first two principal components space). The links between sites are given by the actual distances between the residues (cutoff 10 \AA) on the 3D structure (PDB 1NXW), rescaled as well according to their own value. In d) different colours represent the four sectors (see Supplementary Material).}
\label{fig:networks}
\end{figure*}

Fig.\ref{fig:PF72_lists} a) shows the overlap between the lists of the $n$ most relevant sites according to CVS and SCA. For $n\le 30$ we find that the overlap is smaller than what one would expect from random lists, this meaning that CVS and SCA are sensitive to different statistical information. This is not surprising: SCA is tailored to identify the most relevant drivers of variability in a dataset, and highly conserved sites which, as we saw for our {\it in silico} sequences, are effectively selected by CVS, do not appear on top of the list of relevant sites of SCA. This fact emerges more evidently by the analysis of the single site conservation as a function of the rank of the sites, according to the two methods (see Fig.\ref{fig:PF72_lists} b).  The top sites selected by CVS have indeed much lower entropy than the corresponding ones  for SCA.
SCA and CVS are thus qualitatively very different methods. We can further visualise the difference between the two methods by depicting the $30$ top sites identified by CVS and SCA on the protein tertiary structure. We analysed one structure (PDB ID 1NXW \cite{PFAM}) whose results are shown in Fig. \ref{fig:networks} a) and b). Some common domains on the $\alpha-$helices and $\beta-$ sheets are clearly identified by both methods (red bands). On this structure, a particular site, the Asn-52, has been identified as an active and phosphorylation site \cite{uniprotkb, ncbi-refseq}. Most of the sites identified by only CVS (blue bands) lie indeed around Asn-52 (green star in Fig.\ref{fig:networks}a), whereas those identified only by SCA (purple bands) are much more scattered. 
In particular, the central $\beta-$sheet is singled out by CVS but not by SCA. 

When one compares the lists for larger number $n$ of top sites, the overlap sharply increases. For $n=50$ sites, the overlap between the two lists becomes maximal ($\sim 78\%$) as compared to what one would obtain if sites where  randomly chosen ($\sim 45\%$)\footnote{Since, sequence alignment procedures introduce gaps, we remove those sites that result into gaps in the consensus sequence. The number of selected gaps was very low though, less than the 4\% in both cases.}. Still, the sites identified by CVS happen to be in closer spatial proximity on the three dimensional structure, with respect to those in the SCA list. In order to make this statement more precise, we pictured the two outcomes in a network structure. Each site among the top 50 identified by the two methods defines a node in such a network. The links between two sites are established by their proximity on the 3D structure (PDB ID 1NXW) with a cutoff 10 \AA. The networks for CVS and SCA are represented in Fig.\ref{fig:networks}c) and d). Visual inspection reveals that CVS sites are more densely connected both in terms of number of links and of size of the largest connected component, and less fragmented, meaning that CVS sites are more interacting than those selected by SCA. We also notice that, for SCA, sectors do not seem to be related to a spatial pattern in the network (Fig. \ref{fig:networks}b). Summarising, we can conclude that CVS identifies a \emph{core} of close sites in the protein, localised around the active site Asn-52. 

We also applied SCA to the MSA of the Voltage Sensor Domain. The results are comparable to those previously obtained for the \PF. The overlap between the top $n$ sites of SCA and CVS is smaller than the random threshold for $n<35$, and it increases thereafter, showing a maximum for $n=60$ sites, for which the overlap is about the 80$\%$.

A different way to visualise the difference between the two methods, is to complement our analysis with Direct Coupling Analysis (DCA). DCA is a method aimed at identifying a network of interactions between positions along a protein domain, that are inferred from the traces left by the evolutionary process on the pairwise correlation matrix. In recent years many efforts have been spent in refining this observation into a quantitative bioinformatic tool \cite{weigt2009,morcos2011,cocco2013, ekeberg2013, chris2014, chris2014b, ekeberg2014}. Given the MSA of a certain protein family, DCA usually produces an F-score $F_{i,j}$ for each pair of positions $i,j=1,\ldots,L$ with $F_{i,i}=0$. In particular, if two positions are relevant for preserving the tertiary structure, by establishing a physical contact, one expects residues on these sites to co-evolve, resulting in a large value of $F_{i,j}$. DCA is indeed a powerful tool for predicting contacts in protein domains. 

Here we use DCA to generate a network of interactions between positions, with the goal of deriving an independent assessment of the relevance of the sites selected by CVS and SCA, respectively. We stick with a standard implementation of DCA -- the so called \textit{na\"ive Mean Field Direct Coupling Analysis} (DCA) based on the so-called \textit{Plefka expansion} \cite{plefka1982} of statistical physics that has been shown to capture most of the direct contacts in 3D protein structures \cite{morcos2011} . We refer to the Supplementary Material for a concise discussion of the steps leading to the calculation of $F_{i,j}$ as applied to our dataset. 

\begin{figure}[h!]
\includegraphics[scale=0.4]{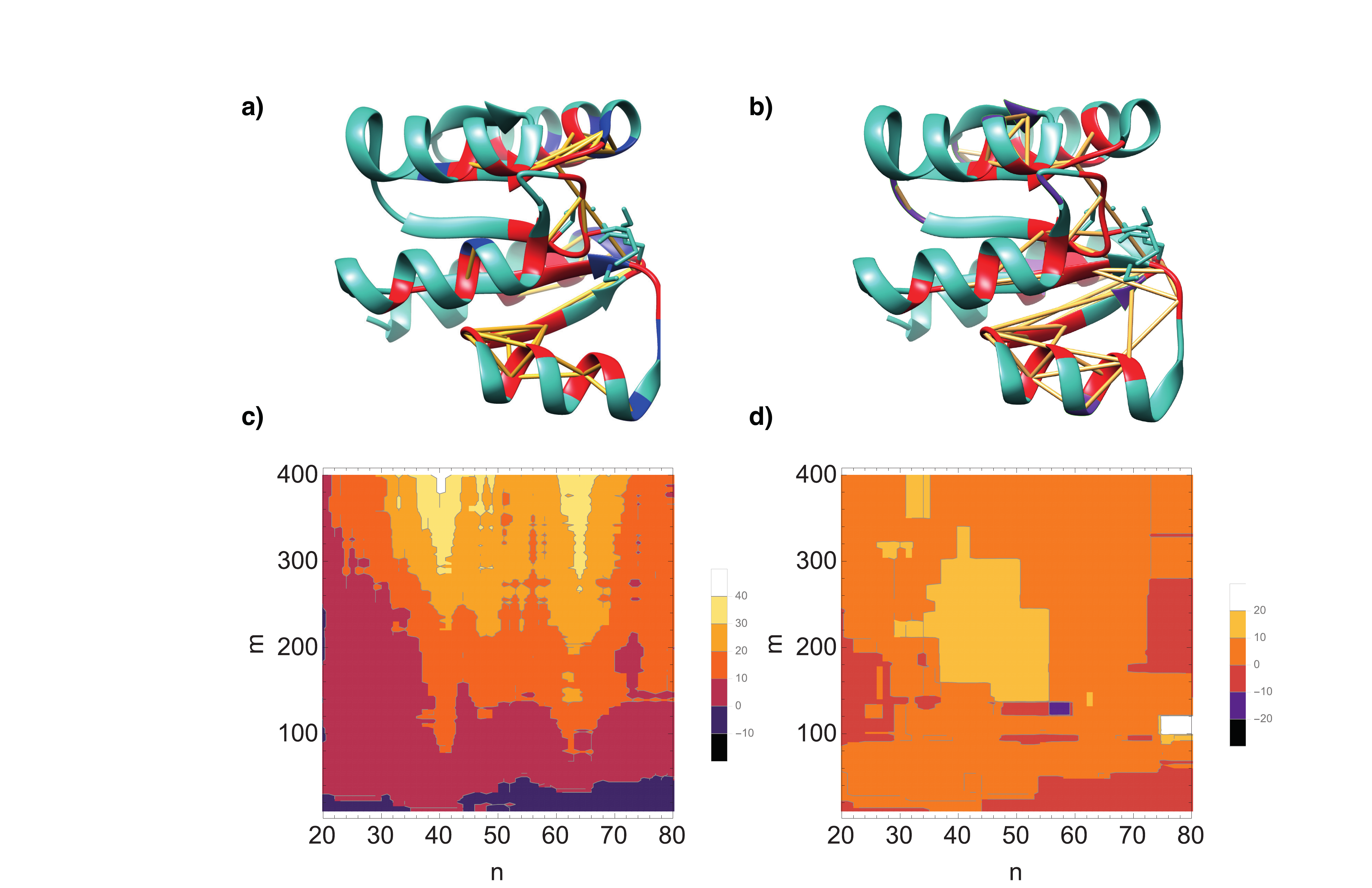}
\caption{a) - b) Top 60 contacts amplified by DCA reduced on the sublists of the top 50 sites highlighted on the 3D structure (PDB 1NXW) for CVS (a) and SCA (b).
c) Density plot of the difference $K_{CVS}-K_{SCA}$ between the number of links in the subnetwork of the $n$ top sites of CVS and SCA, as a function of $n$ and the number of DCA contacts $m$. Here $K_{L}$ is the number of contacts, among the $m$ top contacts identified by DCA, connecting sites belonging to the list $L=CVS,~SCA$. d) Difference between the size of the largest connected component of the sub-networks of the $n$ top sites of CVS and SCA identified by the top $m$ DCA contacts.}
\label{fig:dcanetworks}
\end{figure}

The $m$ contacts with largest value of $F_{i,j}$ define a network among the $n$ top sites of CVS and SCA. A fraction of the $m$ selected links will connect two of the $n$ top sites of a given list, so the density of links and the fragmentation of the resulting networks provides insights on the nature of the co-evolutionary process taking place on the selected sites. Fig. 6 a) and b) displays the links of the $m = 60$ top contacts that connects two of the $n=50$ most relevant sites according to CVS (a) and SCA (b) sites respectively. Interestingly, the putative contacts singled out for CVS turn out to correspond to closer sites with respect to those identified for SCA. In general, sites selected by CVS appear to be more densely connected than those identified by SCA. Indeed, by building up a network using as nodes the top $n$ sites selected by respectively CVS and SCA and as links the top $m$ DCA contacts, one can measure the number of actual contacts, $K_{CVS}$ and $K_{SCA}$, and the size of the largest connected component in the two cases. This allows to notice that that the top $n$ sites identified by CVS share a larger number of the $m$ top DCA contacts and own a larger connected component, with respect to those identified by SCA (see Fig. \ref{fig:dcanetworks} c-d) and Figure caption), in a wide range of values of $n$ and $m$. Furthermore, the number of links between the $n$ top sites and the remaining $L-n$ sites, is smaller for CVS than for SCA (not shown). This is an additional indication of the sharp separation between relevant and irrelevant sites afforded by CVS.

 \section{Biological relevance}
 
 In order to further assess the relevance of sites selected by CVS, we analysed their biological relevance 
comparing them to functional sites stored in databases, the Solvent Accessible Surface rate \cite{naccess1, naccess2} and their importance in the evolutionary dynamics.

\subsection{The response regulator receiver}

We extracted functional sites from the UniProtKB and NCBI-RefSeq databases \cite{uniprotkb, ncbi-refseq} and from \cite{west2001} and checked their rank in the hierarchies defined by CVS and SCA. As shown in Tab. \ref{tab:functionalsitesPF}, most of the functional sites are ranked by both methods among the top 50, showing the ability of both methods of capturing functional features.  Remarkably CVS identifies as top 3 sites two active residues (5,6) and the previously mentioned active and phopshorylation site (50). An important functional domain of this protein, the intermolecular recognition domain, contains one site ranked as 5th by CVS (58) and its surrounding sites are successively ranked,  although captured later on. The same behaviour can be noticed within the dimerisation interface (residues 102,103,104).

\begin{figure}[ht]
\includegraphics[scale=0.6]{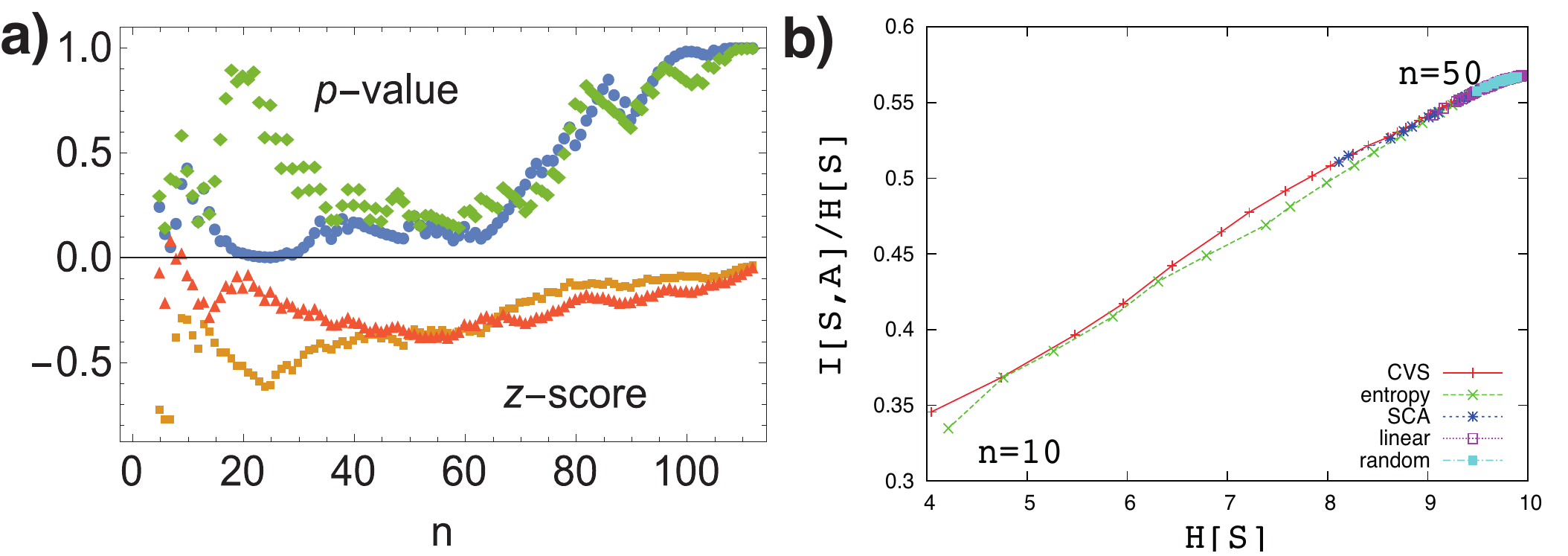}
\caption{a) p-value obtained from the Kolmogorov-Smirnov Test between the Solvent Accessible Surface distributions of the top $n$ relevant sites obtained by CVS (blue circles) and SCA (green diamonds) and the overall one along with the corresponding z-score as defined in the Text for CVS (orange squares) and SCA (red triangles). Solvent Accessible Surface rates have been computed using Naccess \cite{naccess1, naccess2} with PDB 1NXW as input. b) Mutual information between the sequences and the annotations as defined in the Text for different ranking methods (legend shown as inset).}
\label{fig:tabsas}
\end{figure}

Another important biological aspect to investigate concerns the position of the relevant sites on the 3D structure.  To this aim, we analysed how the total CVS counts correlate with the Solvent Accessible Surface rate (SAS) \cite{naccess1, naccess2}. The SAS provides a measure of the accessibility of a given site in the 3D protein structure, to small molecules such as water. Sites with large SAS values are typically on the outer surface of the protein whereas those with low SAS value are buried in the interior. 
We compared the distributions of the SAS of the $n$ most relevant sites identified by CVS and SCA with the overall distributions as well as the z-score of the top $n$ sites identified by the two methods defined as $z-score=\frac{\langle SAS_{x} \rangle_n - \langle SAS \rangle}{\delta SAS}$, where $\langle SAS_x \rangle_n$ is the average SAS of the top $n$ sites identified by $x=CVS, SCA$, $\langle SAS \rangle$ is the average SAS over all the sites and $\delta SAS$ is its standard deviation. Fig.\ref{fig:tabsas} b) shows that the sites selected by CVS have a consistent bias towards lower values of the SAS, indicating that CVS preferably selects internal and conserved sites,  which are putatively important for the maintenance of structural properties of the protein. 
The sites selected by SCA are also preferentially interior, but the bias is considerably weaker.

Finally, as discussed in the Introduction, these sequences all come from a common evolutionary history, being selected as the optimal solution to carry out the same function across different organisms. We here asked then whether the sites identified by CVS can be the carriers of the relevant evolutionary dynamics. To assess this point, we computed the mutual information between the subsequences identified by different methods and the annotation that identifies the organism of origin for each sequence.  We compare the lists of the  top rank $n$ sites obtained according to CVS, SCA and just simple conservation (i.e. the $n$ positions with the smallest site entropy) among themselves and with a list of randomly chosen sites. Different lists afford different levels of variation (i.e. different subsequence entropies $H[\us]$), so a 
meaningful comparison is not at constant $n$ but at constant $H[\us]$.
The fraction of this variation, that accounts for the variation between organisms, was found to be largest for CVS compared to both SCA and conservation, for $n\in [10,50]$ (see Fig. \ref{fig:tabsas}b).

This analysis lets us conclude that CVS is able to extract functional sites with low SAS along with more conserved sites, usually internal in the protein 3D structure, as well as accounting for sequence variation across different organisms.

\subsection{The Voltage-Sensor Domain of the Ion Channels}
\label{subsec:CVSbioresults}
\label{subsec:data}

To further probe the validity of our results, we studied the Voltage-Sensor domain of the Ion Channels sequences. We investigated the biological relevance of CVS relevant sites, i.e., functionality and Solvent Accessible Surface rates. By running our algorithm $R=100$ times for each subsequence length ($n=5,10,15,\ldots,,60$), one is able to build a full counts statistics. On the full dataset of $N=6652$ sequences, we found a dispersion smaller than 10\%. In order to test the stability, we run it also on a subset of $N=666$ sequences for $n=40$ we found an overlap larger of 76\% with the counts obtained for the full dataset.

\begin{figure*}[htbp]
\centering
\includegraphics[scale=0.6]{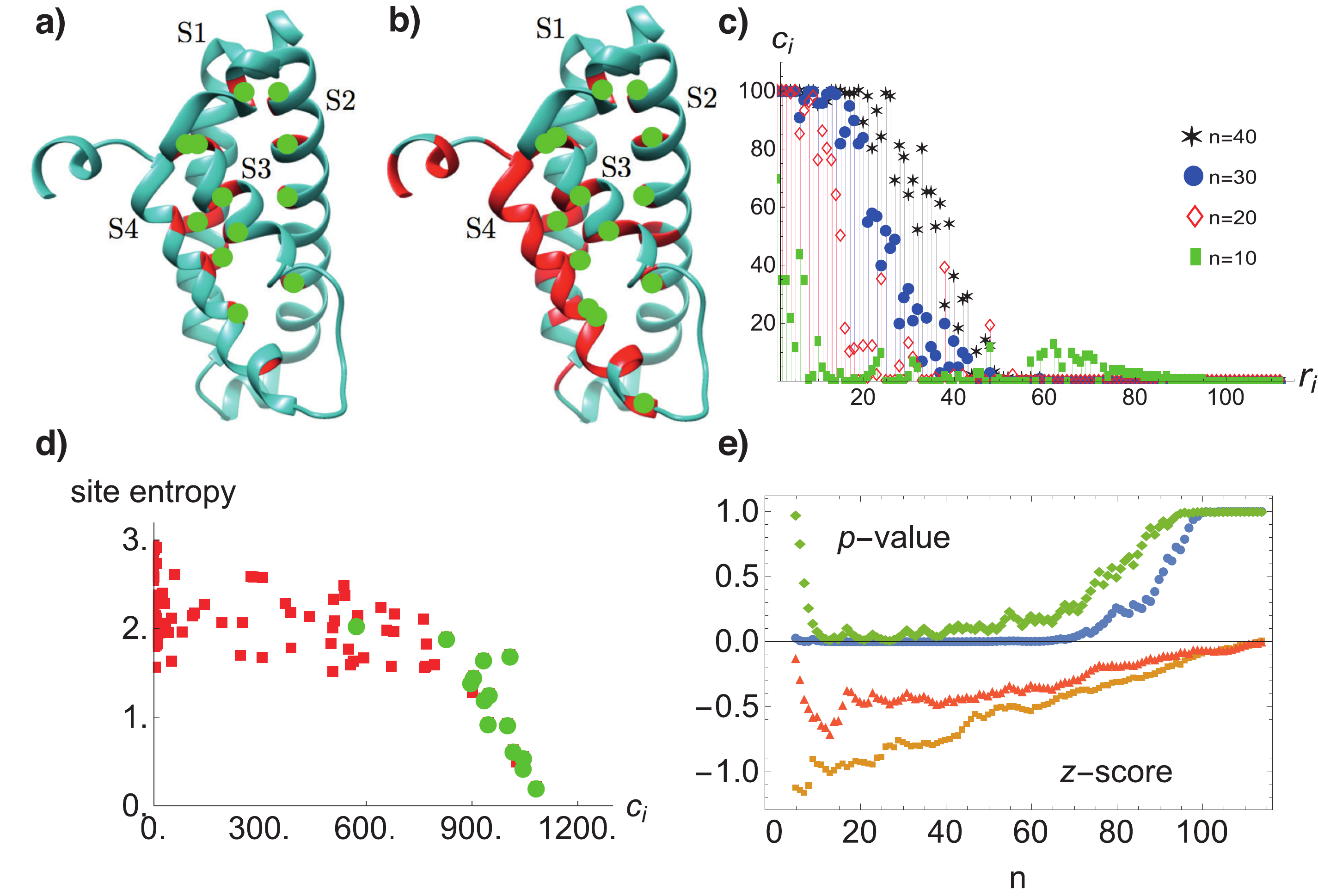}
\caption{a) and b) Top CVS 15 (a) and 40 (b) sites represented on the VSD 3D structure (PDB ID 3RVZ) obtained by running CVS on the dataset 100 times for each subsequence length $n=10,20,30$ and $40$ (respectively, green squares, empty red diamonds, blue circles and black stars) and then ranking sites according to the total count, $C_i$. Green circles spot the functional sites already identified in Refs. \cite{LeePNAS2005, ncbi-refseq} as discussed in the Main Text. c) Ranked relevance count $c_i(n)$ for positions in \MSA for different subsequence lengths, i.e. $n=10,20,30$ and $40$. d) Site entropy as a function of total count. Green circles represent positions identified in the literature (see text) represented on the 3D structure in b) and c). e) p-value obtained from the Kolmogorov-Smirnov Test between the Solvent Accessible Surface distributions of the top $n$ relevant sites obtained by CVS (blue circles) and SCA (red squares) and the overall one. Solvent Accessible Surface rates have been computed using Naccess \cite{naccess1, naccess2} with PDB 3RVZ as input.}
\label{fig:MSA}
\end{figure*}

As shown in Fig. \ref{fig:MSA}d, by ranking the counts, CVS also in this case clearly distinguishes relevant from irrelevant sites affording a sharp division between these two sets. Fig. \ref{fig:MSA}e shows that, besides highly conserved sites, CVS distinguishes between sites whose variability is evolutionarily related from those that can be regarded as noise. By visualising the top ranked sites on the 3D protein structure as well as comparing these results with the available functional knowledge, among these top sites, a first group of 15 positions with high counts $C_i$ can be identified. This contains 9 sites identified in Refs. \cite{MSApaper, LeePNAS2005}. Three more functionally relevant sites have counts larger than 500 belong to a larger group of the 38 most relevant sites. These sites are represented on the 3D structure in Fig. \ref{fig:MSA}b and c. These include N-62, N-72, R-76 and E-93 of 
the voltage-dependent $K^+$ channel KvAP which are important for channel function \cite{LeePNAS2005,ncbi-refseq}. The same sites have also been identified in Ref. \cite{MSApaper}, that refer to the NavAb sequence (E-49, E-59, R-63, D-80 respectively). Ref. \cite{MSApaper} also highlights the role of I-22, F-56 and F-71 in NavAb and it discusses the application of Direct Coupling Analysis on the VSD MSA, identifying several evolutionary conserved contacts along the chain. In particular, E-49 is found to be in contact both with N-25 and with E-96, which are far apart in the NavAb structure. Ref. \cite{MSApaper} argues that these two contacts are important to confer stability both to the activated and to the resting state of the protein domain. All these sites are found to be relevant in the CVS analysis, as well as R-63 and S-77, which are also found to be in contact on the NavAb structure. Ref. \cite{MSApaper} also reports a false positive contact (between W-76 and T-15). We find that while W-76 is relevant, T-15 is not ($C_{T-15}=145$). Finally, we find an enrichment in relevant sites in the region corresponding to S4, which is a highly dynamical region of the VSD, and in the S2-S3 turn (Y-63 to P-95 in KvAP) that has been suggested to be structurally important \cite{LeePNAS2005}. 

We then proceeded again by comparing the Solvent Accessible Surface rate distributions on the top $n$ sites identified by CVS and SCA respectively (see Fig. \ref{fig:MSA} e): in this case, the outcome of the two methods is definitely different up to the top $15$ sites, becoming more similar beyond this threshold, in agreement with what obtained before for the \PF. Again, we find a bias towards internal sites, which is stronger for CVS than for SCA.
The example of VSD sample shows that, in spite of the moderate size of the MSA, CVS is able to extract stable results and to identify functionally relevant sites.

\begin{table*}[!ht] 
\small
\caption{
\ Comparison with databases for CVS and SCA results on \PF. Functional sites have been extracted from \cite{uniprotkb,ncbi-refseq} for the B4DA37$\_$9BACT sequence and \cite{west2001} referring to the structure PDB ID 2CHF. In order to make a consistent comparison sequences have been matched with particular attention to the gaps.}
\label{tab:functionalsitesPF}
\begin{tabular*}{0.7\textwidth}{@{\extracolsep{\fill}}lllll}
\hline
 Source & Our alignment &  Function &  rank SCA &  rank CVS \\  
\hline
\cite{uniprotkb, ncbi-refseq} & 5 & active & 20 & 3 \\ 
\cite{uniprotkb, ncbi-refseq} & 6 & active & 31 & 2 \\ 
 \cite{west2001}     & 47 & binding site for phosphate	 & 83 & 87 \\ 
 \cite{west2001}   & 50 &   phosphorylation     & 36 & 1 \\ 
\cite{uniprotkb, ncbi-refseq} & 52 & active, phosphorylation & 25 & 31 \\ 
\cite{uniprotkb, ncbi-refseq} & 55 & intermolecular recognition site & 78 & 86 \\ 
\cite{uniprotkb, ncbi-refseq} & 56 & intermolecular recognition site &18 & 85 \\ 
\cite{uniprotkb, ncbi-refseq} & 58 & intermolecular recognition site & 35 & 5 \\ 
\cite{uniprotkb, ncbi-refseq} & 59 & intermolecular recognition site & 62 & 84 \\ 
\cite{uniprotkb, ncbi-refseq} & 60 & active, intermolecular recognition site & 59 & 83 \\ 
  \cite{west2001}    & 75 & 	hydrogen bond					       & 98 & 61 \\ 
\cite{uniprotkb, ncbi-refseq} & 80 & active & 44 & 37\\ 
 \cite{west2001} &	97 & 	phosphorylation	& 1 & 14 \\ 
\cite{uniprotkb, ncbi-refseq} & 99 & active & 48 & 62 \\ 
\cite{uniprotkb, ncbi-refseq} & 102 & active and dimerization interface & 8 & 25\\ 
\cite{uniprotkb, ncbi-refseq} & 103 & active and dimerization interface  & 75 & 43 \\ 
\cite{uniprotkb, ncbi-refseq} & 104 & active and dimerization interface & 64 & 46 \\ 
\hline
\end{tabular*}
\end{table*}

\section{Conclusions}

In summary, we have proposed a new method for the identification of a core of functionally and structurally relevant sites in protein domains. Given a multiple sequence alignment (MSA) for a protein family, the method is based on finding those subsets of $n$ positions for which the entropy $H[K]$ of the frequency with which different subsequences occur is maximal, corresponding to broad frequency distributions for these subsequences. Its implementation is straightforward and it does not require any further data-processing step. By starting with different subsequences length, the method assigns to each position a \textit{count} that is used to assess the relevance of a certain site. As the subset length includes enough positions, CVS affords a sharp separation between relevant and irrelevant sites for all the datasets we analysed, the relevant sites are often but not always highly conserved positions. Besides, site relevance turns out to increase with the subset size $n$, i.e., typically, if a site $i$ is selected as relevant in the subset of the $n$ most relevant positions, it will be very likely selected in the subset of $n'>n$ sites as well. 

The application of our method to \emph{in silico} sequences provides a first check of the ability of CVS of discriminating relevant information from noise and capturing dependencies going beyond pairwise correlation in big datasets. 

We have discussed the application of the method to two protein domain families, the response regulator receivers (RR, PFAM ID PF00072) and the voltage sensor domain of the ion channels (VSD, PFAM ID PF00520). We first studied the response regulator receivers (\PF) and inspected the robustness of the method against reweighing and its ability of going beyond pure single site conservation and pairwise correlations. We then compared our method with Statistical Coupling Analysis (SCA) by analysing the RR and the VSD. After assigning a measure of relevance to both methods, we studied the overlap between the two solutions finding out that although the top most relevant sites are normally quite different, this overlap increases. For the top 50 sites in the RR, the overlap between the two solutions is 78$\%$. An analogous result was found for the VSD. Yet, the small differences between CVS and SCA results can be furthermore highlighted by using Direct Coupling Analysis (DCA). CVS identifies a core of densely connected residues and all significant contacts predicted by DCA on the restricted lists turn out to be close on the 3D structure for the \PF. 
CVS is furthermore able to identify biologically relevant positions: the sites extracted from \cite{uniprotkb, LeePNAS2005,ncbi-refseq} or identified by inference methods as Direct Coupling Analysis turn out to be tagged as relevant by our method for both RR and VSD. 
This further corroborates the conclusion that CVS indeed singles out subsets of relevant sites in protein domains.

We stress, in particular, the fact that CVS is able to recover insights from methods based on  single site conservation and pairwise correlation. Yet, the most exciting aspect of CVS lies precisely in its ability to probe the co-evolutionary process beyond single site conservation and pairwise correlation. This calls, on one hand, for the development of inference methods going beyond pairwise interactions, and on the other hand to applications of CVS to instances that may lead to a more critical assessment of its potential for reverse engineering evolutionary processes. 

\section*{Acknowledgments}

We thank S\'ebastien Boyer, Alice Coucke and Eleonora De Leonardis for helpful suggestions, 
Vincenzo Carnevale and Daniele Granata for interesting discussions and for providing the data on \MSA, 
Thierry Mora, Olivier Rivoire, Yasser Roudi and Pierpaolo Vivo for insightful discussions and Christoph Feinauer for providing us with the pairwise-correlations constrained reshuffled dataset. We would like to particularly acknowledge Matteo Figliuzzi and Martin Weigt for their critical reading of the manuscript and their enriching suggestions on this work. 
This work was supported by the Marie Curie Training Network NETADIS (FP7, 290038). The codes for the CVS algorithm and the analysis done in this paper are available on request.

\bibliographystyle{unsrt}
\bibliography{revisionJan.bbl}

\renewcommand\appendixname{Supplementary Material}
\appendix
\title{Supplementary Material to \\ 
Identifying relevant positions in proteins by Critical Variable Selection}
\author{Silvia Grigolon, Silvio Franz, Matteo Marsili}

\setcounter{figure}{0} \renewcommand{\thefigure}{S\arabic{figure}}


%
\section{Statistical Coupling Analysis in a nutshell}
\label{sec:sca}

We here give an overview of Statistical Coupling Analysis as applied to our datasets. 

Let us consider a MSA as an ensemble of $N$ sequences $\vec{s}^\alpha=\{s^\alpha_1,...,s^\alpha_L\}$ of length $L$ where each $s^\alpha_i$ ($\alpha=1,...,N$ and $i=1,...,L$) represents either an amino acid or a gap or an uncertain letter as well and can then take $q=21$ values. A first measure of conservation throughout the dataset is given by the frequency of the amino acid $a$ at position $i$, i.e.:
\begin{equation*}
f^a_i \equiv \frac{1}{N} \sum_{\alpha=1}^{N} \delta_{a,s^{\alpha}_i}.
\end{equation*}
Pair - frequencies can be also defined in a straightforward manner, as:
\begin{equation*}
f^{ab}_{ij} \equiv \frac{1}{N} \sum_{\alpha=1}^{N} \delta_{a,s^{\alpha}_i} \delta_{b,s^{\alpha}_j},
\end{equation*}
that gives a measure of the simultaneous appearance of the amino acids $a$ and $b$ respectively at positions $i$ and $j$. 
The \textit{correlation matrix $C^{ab}_{ij}$} in such defined model will be then:
\begin{equation}
C^{ab}_{ij} \equiv f^{ab}_{ij}-f^a_i f^b_j,
\end{equation}
being a $q L \times q L$ matrix. In \cite{halabi2009}, a further quantity, $\phi_i^a$, called \textit{positional information} was  introduced. This is aimed at highlighting highly conserved positions with respect to the background amino acids frequencies within the correlation matrix. Let us define the background frequency of the $a-$th amino acid as $\nu^a=\frac{1}{L} \sum_{i=1}^{L} f_i^a$. The bias of a site $i$ towards one particular amino acid with respect to the background can be quantified  by the Kullback - Leibler divergence, $D_{KL}(f_i || \nu)=\sum_{a=1}^q f_i^a \quad \mbox{log}(\frac{f_i^a}{\nu^a})$. The {\it positional information} is defined as $\phi_i^a \equiv \frac{\partial D_{KL}(f_i || \nu)}{\partial f_i^a}$. 

In \cite{halabi2009} it was suggested to rescale 
the correlation matrix $C^{ab}_{ij}$ taking into account the 
positional information as follows:
\begin{equation}
\tilde{C}^{ab}_{ij} = (f^{ab}_{ij}-f^a_i f^b_j) \phi_i^a \phi_j^b.
\end{equation}

In order to avoid singularities due to the presence of the logarithm in the Kullback-Leibler divergence, we used pseudo counts to regularise frequencies \cite{morcos2011, lunt2010}, i.e., adding at each position a fictive count.  

In addition, due to sampling biases \cite{sequencing1, mardis2011, pagani2012}, the dataset is not spatiotemporally homogenous and with an overabundance of some specific very similar sequences. To limit this bias, we have reweighed 
sequences by collapsing those overlapping at least of the 90$\%$, which we will refer to as \textit{similarity threshold} $\sigma$. We verified that  values $0.9 \leq \sigma \leq 1$ does not change sensibly the results for the families we analysed. In the following we shall call the number of effective sequences $M_{eff}$. 

Regularisation and reweighing can be expressed in a compact manner for single-site and pair frequencies as follows:

\begin{equation}
f^a_i=\frac{1}{M_{eff}+\lambda M_{eff}} \Big( \frac{\lambda}{q}+\sum_{\alpha=1}^{M_{eff}} \delta_{a,s^\alpha_i} \Big)
\label{eq:pseudosingle}
\end{equation} 

and 

\begin{equation}
f^{ab}_{ij}=\frac{1}{M_{eff}+\lambda M_{eff}} \Big[\frac{\lambda}{q}\delta_{ij}\delta_{ab}+\frac{\lambda}{q^2}(1-\delta_{ij}) + \sum_{\alpha=1}^N \delta_{a,s^\alpha_i} \delta_{b,s^\alpha_j}\Big],
\end{equation}

where $\lambda=1$ is the pseudo count. 

The regularised and reweighed $\tilde{C}^{ab}_{ij}$ is still a $qL \times qL$ matrix: to reduce it to a $L \times L$ matrix, we used the so-called \textit{Frobenius norm}, i.e.:
\begin{equation}
\bar{C}_{ij}=\sqrt{\sum_{a,b=1}^q {\tilde{C}_{ij}^{ab \mbox{  } 2}}},
\end{equation}

$\bar{C}_{ij}$ is now a $L \times L$ symmetric matrix. Such a matrix does not show the usual properties of a typical correlation matrix: its diagonal elements are indeed not unitary and this is due to the rescaling procedure aimed at highlighting the conservation at each position. Note that this reflects the main aim of the original SCA, i.e., to take into account at the same time both pairwise correlations and positional conservation. To perform SCA, as we previously discussed, one must compare the spectral properties (i.e., eigenvalues and eigenvectors' components) of the $\bar{C}_{ij}$ with those of the correlation matrix got from the reshuffled dataset. Data reshuffling is performed constraining on the single amino acid frequency at each site, i.e., randomly exchanging two different amino acids at the same position $i$. The procedure to compute the correlation matrix is exactly the same as before and we call the random matrix $\mathcal{M}_{ij}$. 

\begin{figure}[h!]
\includegraphics[scale=0.7]{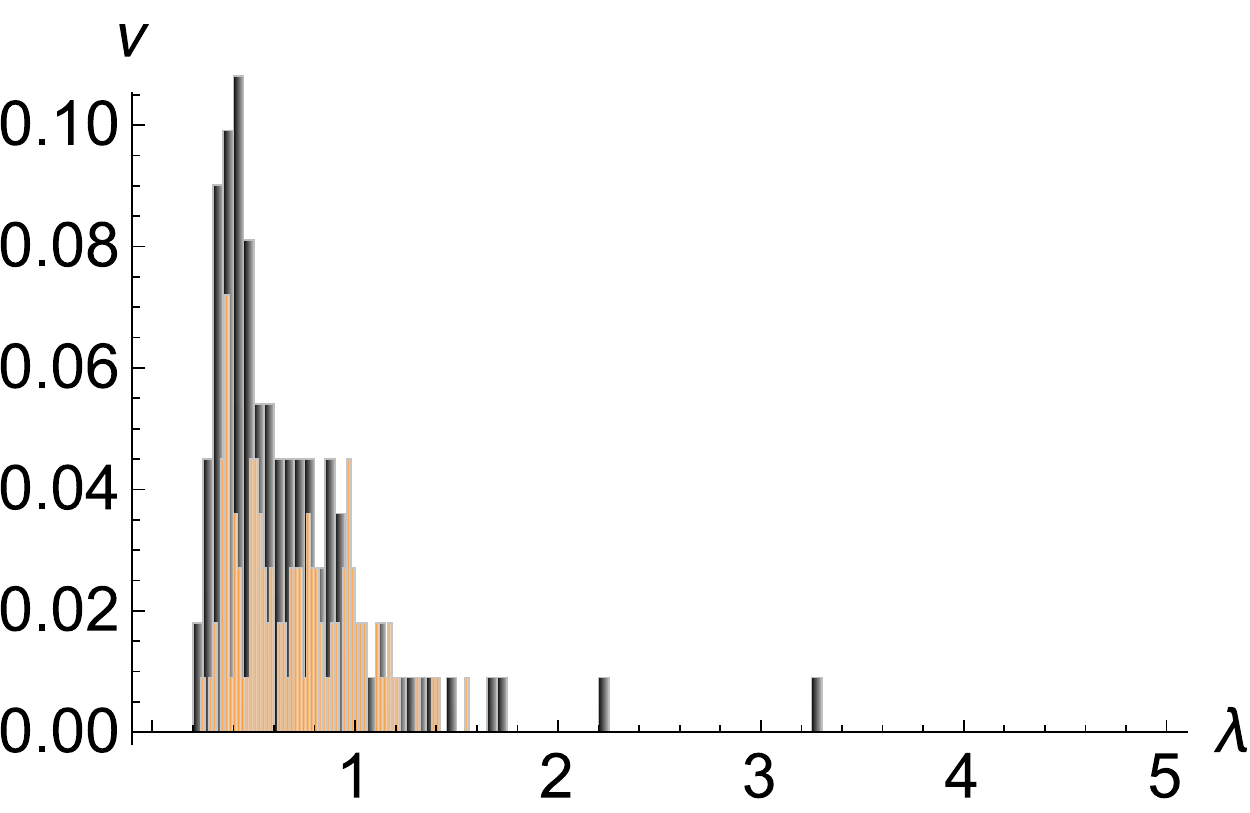}
\caption{Eigenvalues distribution for the actual correlation matrix (orange blocks) and the random matrix obtained from the reshuffled dataset (black blocks).}
\label{fig:eigenvaluesPF}
\end{figure}

\begin{figure}
\includegraphics[scale=0.4]{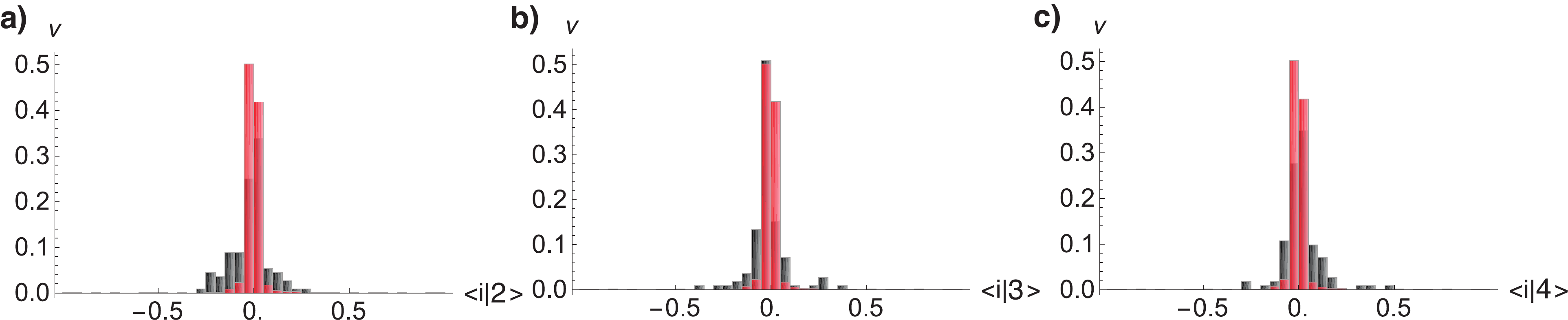}
\caption{Histograms of eigenvectors components frequencies (second eigenvectors in a), third in b) and fourth in c)), $\nu$, relatively to the correlation matrix $\bar{C}_{ij}$ (black blocks) and to the random one (red blocks).}
\label{fig:eigenmodes}
\end{figure}

\begin{figure}
\includegraphics[scale=0.4]{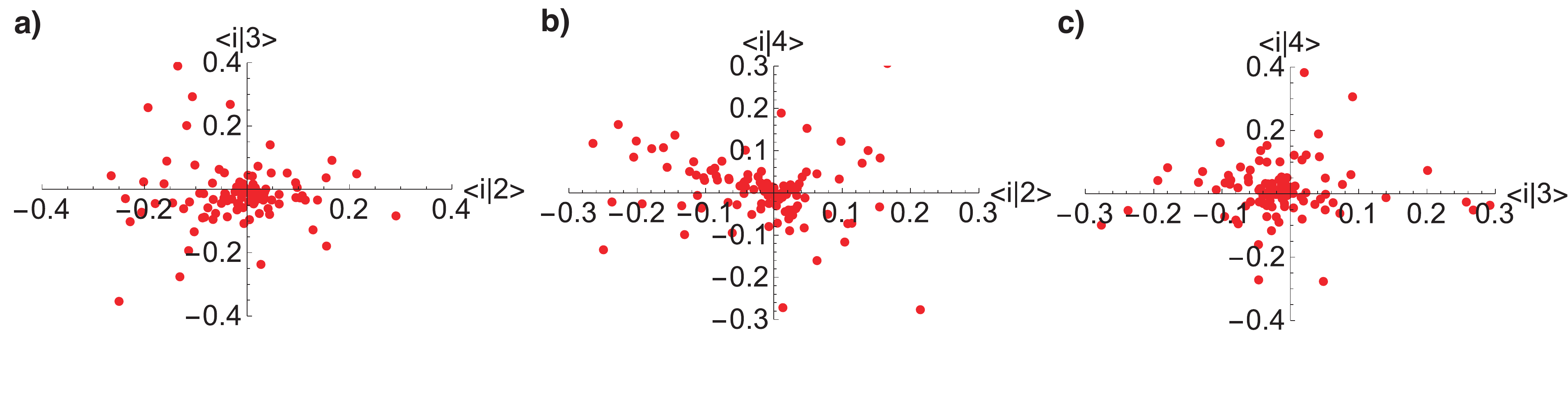}
\caption{Eigenvectors components for the matrix $\bar{C}_{ij}$. While in a) one can still see a clusters structure, in the others the eigenvectors' components are much noisier.}
\label{fig:eigencomparisons}
\end{figure}

As introduced in the Main Text, to figure out whether some relevant information is enclosed in the dataset, one has firstly to compare eigenvalues' distributions relatively to the $\bar{C}_{ij}$ with those of $\mathcal{M}_{ij}$. We expect $\mathcal{M}_{ij}$'s eigenvalues distribution to be Marchenko-Pastur like \cite{marchenko1967}, i.e., a bulk of very small eigenvalues and short tails. Fig. \ref{fig:eigenvaluesPF} shows at least four eigenvalues (black blocks) of the $\bar{C}_{ij}$ computed for the \PF rising out of the random bulk (orange blocks). The first highest eigenvalue has not been shown since it is a consequence of the phylogenetic history characterising the dataset  \cite{halabi2009} and of the use of pseudo counts and it will not be taken into account for sectors selection. 
In order to identify \textit{sectors}, we sticked with the second and third highest eigenvalues, $\lambda_2$ and $\lambda_3$ (Fig. \ref{fig:eigenvaluesPF}) and their associated eigenvectors, $| 2 \rangle$ and $| 3 \rangle$ (Fig.\ref{fig:eigenmodes}). The aim is to display those sites giving \textit{signals} along these directions, i.e., having a projection along the two eigenvectors significantly higher than the random one. Commonly, one defines a discrimination threshold $\epsilon$ to distinguish the randomness in the eigenvectors' components from actual biologically relevant signals \cite{halabi2009}. As our aim consists of mainly comparing the relevant sites identified by CVS with those identified by SCA, we first introduced a measure of relevance in SCA as well. Let us consider the eigenvectors associated with the second and third highest eigenvalues of the correlation matrix $\bar{C_{ij}}$ and the projection of the correlation matrix along these directions (Fig.\ref{fig:eigencomparisons}). As shown in Fig.\ref{fig:eigenmodes}, most of the randomness will be localised around small values of the eigenvectors' components. In turn, actual relevant signals can be detected far from this random bulk. We thus defined the relevance of the $i-th$ position as the distance of the $i-th$ point in the plane spanned by the eigenvectors associated to the previously mentioned highest eigenvalues. The most relevant sites will be then the most far from the origin of this plane. 
To compare CVS results to those obtained by SCA, we then sorted both lists of sites according to the relevance definition in each of these methods and we computed the overlap between the two lists by considering the top $n$ positions. As discussed in the Main Text, for small values of $n$ the overlap is very small, thus the two methods actually give very different results. Increasing $n$, the overlap increases and becomes quite different from the random one, till reaching a maximum for values of $n \simeq L/2$ (Fig.4a). We thus chose this value of $n$ for our comparisons shown in the Main Text.




\begin{figure}
\includegraphics[scale=0.5]{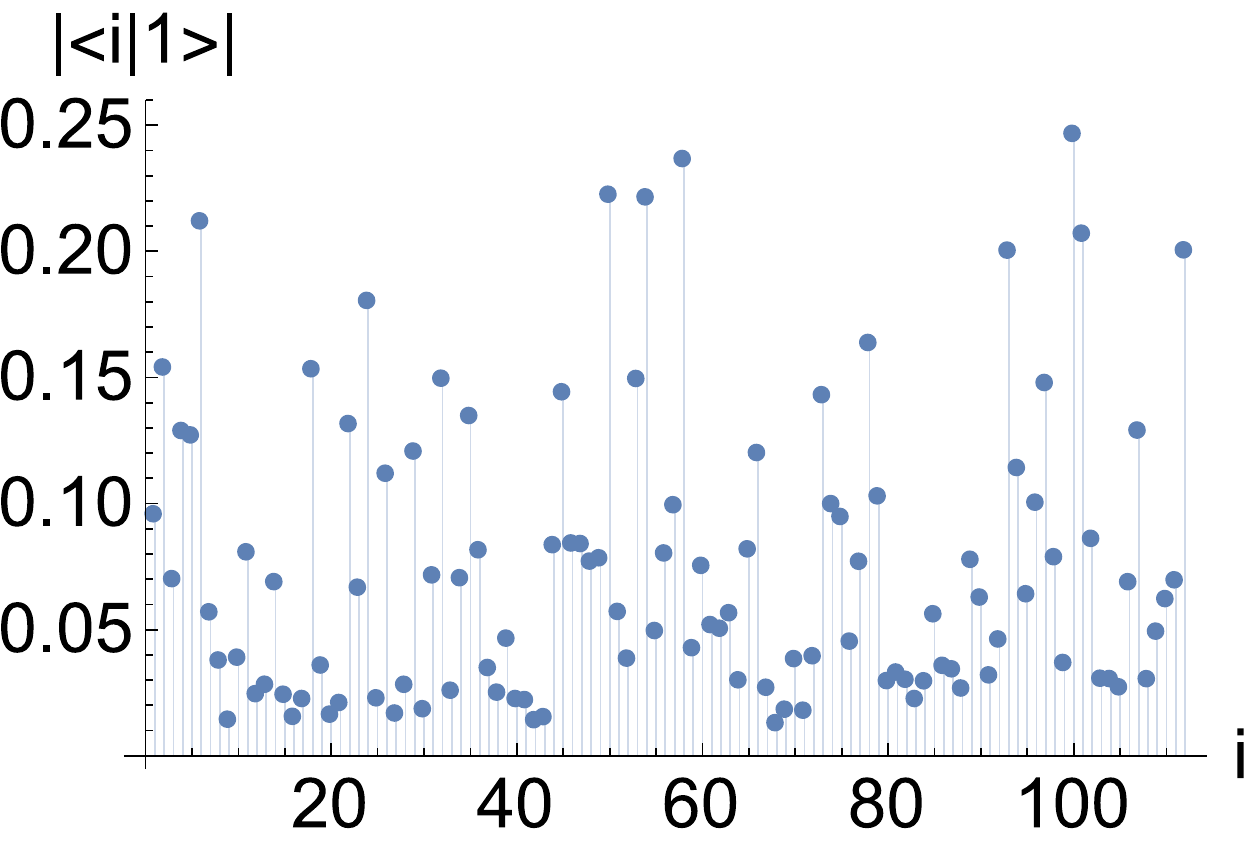}
\caption{Absolute value of the first eigenvector components, $|\langle i | 1 \rangle|$, plotted along the sequence. Notice that most of them are at least higher than 0.025.}
\label{fig:firstcomponentPF}
\end{figure}

\begin{figure}
\includegraphics[scale=0.6]{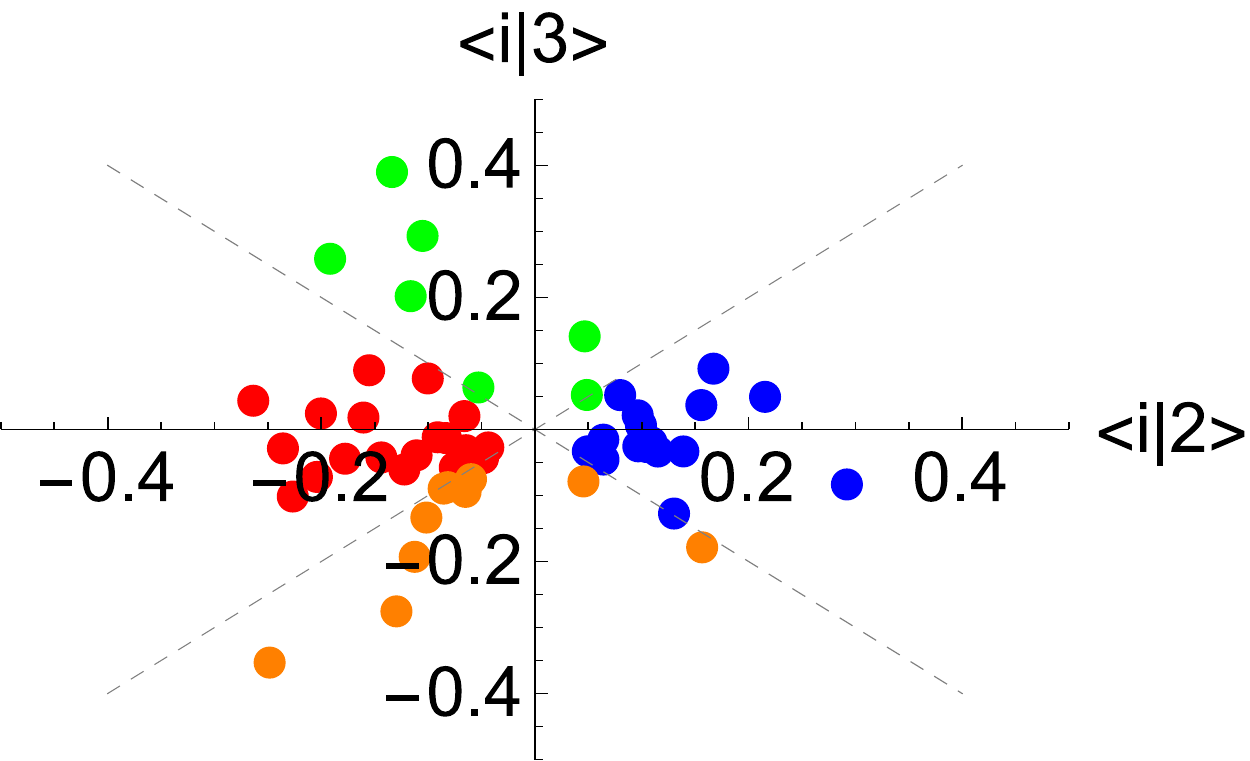}
\caption{Sectors identified by clustering the positions along the protein in the plane spanned by the principal components. The clustering procedure has been performed as explained in the Text.}
\label{fig:PF72sectors}
\end{figure}

For completeness, we also defined sectors for the SCA results, following the same clustering procedure as in \cite{halabi2009}. Note however that here no discrimination threshold is imposed but sites are sorted according to their distance in the plane spanned by the principal components. 
For \PF, four sectors have been identified by grouping the positions in the plane spanned by $| 3 \rangle$ and $| 2 \rangle$ in the following way:
\begin{itemize}
\item the first sector is identified by all those positions $\langle i | 2 \rangle > 0$ and $| \langle i | 2 \rangle | > | \langle i | 3 \rangle |$;
\item the second sector is identified by all those positions $\langle i | 2 \rangle < 0$ and $| \langle i | 2 \rangle | > | \langle i | 3 \rangle |$;
\item the third sector is identified by all those positions $\langle i | 3 \rangle > 0$ and $| \langle i | 2 \rangle | < | \langle i | 3 \rangle |$;
\item the fourth sector is identified by all those positions $\langle i | 3 \rangle < 0$ and $| \langle i | 2 \rangle | < | \langle i | 3 \rangle |$.
\end{itemize}

The obtained sectors are shown in Fig. \ref{fig:PF72sectors} and their meaning is discussed in the Main Text. 

Fig. \ref{fig:PF72sectors} plots the first 50 points in this list for \PF in the space spanned by the 2nd and 3rd principal components (as discussed in \cite{halabi2009}, the largest eigenvalue should not be considered since it is a signature of the phylogenetic history of the dataset). Performing a clustering procedure onto this set, one finally gets groups of mostly correlated sites, usually called \textit{sectors} in the literature \cite{halabi2009,cocco2013}. 
For \PF, four sectors can be identified, corresponding to functional domains on the tertiary structure. Within CVS instead no notion of sectors has been defined yet: we thus stick with the only notion of relevance given by the counts discussed in Main Text.

In our analysis we just sticked with the two highest eigenvalues of the correlation matrix. Yet, as pointed out in \cite{cocco2013}, smaller eigenvalues can still give signals about some positions. However, we found that CVS recovers most of these positions found to be relevant along eigenvectors associated to smaller eigenvalues. 

\section{Direct Coupling Analysis in a nutshell}

Direct contact prediction from MSAs has recently been subject of intense research. Here we focus on the well-known method of Direct Coupling Analysis. As discussed in the Introduction of Main Text, many approaches have been proposed so far some aimed at minimising the detection of false positives while some others at weighing the gaps introduced by the alignments. Hereafter, we will refer to \cite{morcos2011} where they introduced and tested the so-called naive Mean Field  approach (nMFDCA) to infer the interactions between the amino acids. 

The ansatz of DCA methods is that each sequence is the outcome of a Boltzmann - distribution, $P(\vec s)$, obtained from a maximum entropy principle under the constraints that marginal distributions must match the experimental ones, i.e.:
\begin{center}
$P(s_i=a) \equiv f^a_i$\\
and\\
$P(s_i=a, s_j=b) \equiv f^{ab}_{ij}$.
\end{center}
This allows the introduction of a q-state Potts' hamiltonian, $\mathcal{H}$, given by:
\begin{equation}
\mathcal{H}[\vec s]=-\sum_{i < j} J_{ij}(s_i, s_j)- \sum_{i} h_i({s_i}).
\end{equation}
The model we are aimed at fitting data with is then a 21-state Potts' model.

Since for each site $i$ frequencies sum up to 1, being there L constraints for each site, the model has $(q-1) L$ free parameters which can be inferred exploiting the Plefka's expansion of the Gibbs' free energy generalized to the q-state Potts' model: it relates couplings $J_{ij}(s_i,s_j)$ to the $(q-1) L \times (q-1) L$ correlation matrix $C_{ij}(s_i,s_j)$ \cite{plefka1982, morcos2011}. The correlation matrix $C_{ij}(s_i,s_j)$ is defined as in SCA other than the rescaling with positional conservation \cite{rivoire2013}. 
One can check that from the Plefka's expansion it follows that:
\begin{equation}
J_{ij}({s_i, s_j})=-(C_{ij}({s_i, s_j}))^{-1}.
\end{equation}

However, to ensure matrix inversion, one must neglect the $q-$th degree-of-freedom because of the $L$ frequency constraints we discussed before. Here, the use of pseudo counts is fundamental in order to avoid singularities due to positional under sampling. 

This allows to obtain a regular $J_{ij}({s_i, s_j})$ matrix whose dimensions are $(q-1)L\times(q-1)L$: to perform a dimensional reduction on the couplings and turning again to a $L \times L$ matrix, one can introduce again the $q-th$ degree-of-freedom (as a null column/row) and then to standardize the couplings in the following way:
\begin{equation}
\tilde{J}_{ij}({s_i, s_j})=J_{ij}({s_i, s_j})-\mu_{ij}({s_i})-\mu_{ij}({s_j})+\mu_{ij},
\end{equation}
where $\mu_{ij}({s_i})=\frac{1}{L}\sum_{s_j=1}^q J({s_i, s_j})_{ij}$ (analogously for $\mu_{ij}({s_j})$) and $\mu_{ij}=\frac{1}{L^2}\sum_{s_i,s_j=1}^q J_{ij}({s_i, s_j})$ and then take the Frobenius norm as previously defined. 
The Frobenius norm computed on this new couplings matrix is called the \textit{$F$--score}, defined as:
\begin{equation}
F_{ij} \equiv || \tilde{J}_{ij}({s_i, s_j}) ||_{s_i , s_j} = \sqrt{\sum_{s_i, s_j=1}^q {J_{ij}({s_i, s_j})}^2}.
\end{equation}
The F-score turns out to have zero elements on the diagonal, i.e., zero self-couplings, and a better highlight of structures within the coupling matrix \cite{cocco2013}. 

\end{document}